\documentclass[a4paper,fleqn,usenatbib]{mnras}
\usepackage{graphicx}	
\usepackage{amsmath}	
\usepackage{amssymb}	
\usepackage{comment,soul,orcidlink}

\def\lsun{{\rm L_{\odot}}}
\def\msun{{\rm M_{\odot}}}
\newcommand\rsun{{\rm R_{\odot}}}

\def\be{\begin{equation}}
\def\ee{\end{equation}}

\def\del#1{{}}

\newcommand\mj{{\,{\rm M}_{\rm J}}}

\newcommand{\bref}[1]{{ #1}}

\newcommand\MSunPerYear{~${\rm M_{\odot}}$~yr$^{-1}$\,}

\newcommand\konkoly{Konkoly Observatory, Research Centre for Astronomy and Earth Sciences,\\Eötvös Loránd Research Network (ELKH), Konkoly-Thege Mikl\'os \'ut 15--17, 1121 Budapest, Hungary}
\newcommand\elte{ELTE E\"otv\"os Lor\'and University, Institute of Physics, P\'azm\'any P\'eter s\'et\'any 1/A, \\ 1117 Budapest, Hungary}
\newcommand\heidelberg{Max Planck Institute for Astronomy, Königstuhl 17, 69117 Heidelberg, Germany}

\title[Thermal Instability and episodic accretion]{Episodic eruptions of young accreting stars: the key role of disc thermal instability due to Hydrogen ionisation.}

\author[Nayakshin et al.]{Sergei Nayakshin$^{1 \orcidlink{0000-0002-6166-2206}}$\thanks{sergei.nayakshin@le.ac.uk}, Fernando Cruz S\'aenz de Miera$^{2,3,4 \orcidlink{0000-0002-4283-2185}}$, 
\'Agnes K\'osp\'al$^{3,4,5,6 \orcidlink{0000-0001-7157-6275}}$,  
\newauthor
Aleksandra \'{C}alovi\'{c}$^{1}$,
Jochen Eisl\"offel$^{7}$ and Douglas N.C. Lin$^{8,9}$
\\
$^{1}$School of Physics and Astronomy, University of
  Leicester, Leicester, LE1 7RH, UK. \\
$^2$ Institut de Recherche en Astrophysique et Planétologie, Université de
Toulouse, UT3-PS, CNRS, CNES, 9 av. du Colonel Roche, \\
31028 Toulouse Cedex 4, France \\
$^3$\konkoly\\
$^4$ CSFK, MTA Centre of Excellence, Konkoly-Thege Miklós út 15-17, 1121 Budapest, Hungary\\
$^5$ \elte\\
$^6$ \heidelberg\\
$^7$ Thüringer Landessternwarte Tautenburg, Sternwarte 5, 07778 Tautenburg, Germany\\
$^8$ Department of Astronomy \& Astrophysics, University of California, Santa Cruz, CA 95064, USA\\
$^9$ Institute for Advanced Studies, Tsinghua University, Being 100084, China
}

\date{Accepted XXX. Received YYY; in original form ZZZ}

\pubyear{2023}

\begin{document}
\label{firstpage}
\pagerange{\pageref{firstpage}--\pageref{lastpage}}
\maketitle

\begin{abstract}
In the classical grouping of large magnitude episodic variability of young accreting stars, FUORs outshine their stars by a factor of $\sim 100$, and can last for up to centuries; EXORs are dimmer, and last months to a year. A disc Hydrogen ionisation Thermal Instability (TI) scenario was previously proposed for FUORs but required \bref{unrealistically low} disc viscosity. In the last decade, many intermediate type objects, e.g., FUOR-like in luminosity and spectra but EXOR-like in duration were found.  Here we show that the intermediate type bursters Gaia20eae, PTF14jg, Gaia19bey and Gaia21bty \bref{may be naturally explained} by the TI scenario with realistic viscosity values.  We argue that TI predicts a dearth (desert) of bursts with \bref{peak} accretion rates between $10^{-6}$\MSunPerYear  $\lesssim \dot M_{\rm burst} \lesssim 10^{-5}$\MSunPerYear, and that this desert is seen in the sample of all the bursters with previously determined $\dot M_{\rm burst}$. Most classic EXORs (FUORs) appear to be on the cold (hot) branch of the S-curve during the peak light of their eruptions; \bref{thus TI may play a role in this class differentiation}. 
At the same time, TI is unable to explain how classic FUORs can last for up to centuries, \bref{and over-predicts the occurrence rate of short FUORs by at least an order of magnitude}. We conclude that TI is a required ingredient of episodic accretion operating at $R\lesssim 0.1$~au, but additional physics must play a role at larger scales. \bref{Knowledge of TI inner workings from related disciplines may enable its use as a tool to constrain the nature of this additional physics.}

\end{abstract}

\begin{keywords}
stars:formation -- stars: protostars -- protoplanetary discs 
\end{keywords}

\section{Introduction}


While the steady-state accretion disc solution of \cite{Shakura73} applies in many circumstances, it is well known that accretion discs can be thermally and viscously unstable \citep{LightmanEardley-74,Pringle76,BathPringle82-S-curve}. One of such situations arises when the steady-state solution predicts that there is a radius in the disc where Hydrogen should suddenly transition from being neutral to ionised. Instead of a discontinuous transition at this radius, real discs develop time-dependent oscillations in which the inner disc keeps switching  between the cold neutral and the hot ionised states \citep{Meyer84-CVs}. For a simple but insightful overview of the subject see \S 6.1.1 in \cite{Armitage15-review}. This theory has been successfully applied to unsteady accretion outbursts in Dwarf Novae (DNe) and Low mass X-ray binaries \citep[e.g.,][LMXRBs]{Smak84-dwarf-novae-review,Cannizzo93-DNe,Lasota01-Review}. One of the inescapable conclusions of this work is the realisation that the $\alpha$-viscosity parameter needs to be $\alpha_{\rm c} \sim (0.01-0.03)$ at the low (cold) stable branch of the S-curve and about a factor 10 larger at the high (hot) stable branch \citep[e.g.,][]{Smak84-alpha,Hameury98-alpha}. Recent 3D radiation and magnetohydrodynamics simulations self-consistently arrived at such values of viscosity \citep{Hirose14-Convection,Hirose15,Scepi-18-convection,Scepi-18-TI-alpha}. While important details of the outbursts are not yet fully understood \citep{Coleman16-TI,Scepi-19-DNe}, there can be no doubt that TI is the correct first-order framework for interpreting large amplitude outbursts in DNe and LMXRBs \citep{Hameury-20-review}. \bref{For example, \cite{Coriat12_DIM_LMXRB} and \cite{Dubus18_DIM_DNe} collected samples of $\sim 50$ and $\sim 130$ LMXRBs and DNe, respectively, known at the time. Unlike protoplanetary discs, discs in these systems have very small radial extent (typically smaller than $\rsun$!) that depends on binary separation. It is thus possible to predict for which accretion rates these systems should be accreting stably because the whole of the disc is on the hot branch of the S-curves. There is also a well established accretion rate for which all of the disc is on the cold branch. Systems with intermediate values of accretion rate should be unstable. \cite{Coriat12_DIM_LMXRB} and \cite{Dubus18_DIM_DNe} found that their respective samples are in excellent agreement with this theoretical prediction.}

During the earliest phases of protoplanetary disc evolution, their inner regions are viscously rather than passively heated \citep[e.g.,][]{Hartmann16-review}, and so their physics should be somewhat similar to that of the discs in DNe and LMXRBs. Protoplanetary discs may therefore display disc thermal instabilities too at sufficiently high accretion rates \citep{HartmannK85-FUORs}.
Observations of the last century indeed showed that young accreting protostars experience 
large magnitude episodic variability that can be classified into two main classes: FUORs, objects experiencing optical brightening by up to 6 magnitudes that last for $\sim$ centuries, and EXORs, which show less dramatic increases in luminosity, and outburst duration of order months to a year \citep[e.g.,][]{Herbig89,HK96,Clarke05-FUORs}. There are important spectroscopic differences between the two classes. FUOR spectra are clearly dominated by emission from a self-luminous accretion disc, whereas EXORs spectra are dominated by the star, accretion shocks on its surface, and radiation reprocessed by passive (irradiated) discs \citep[e.g.,][]{Connelley18-FUORs,Liu22-FUORs,Wang-23-EXLUP}. Unsurprisingly, the  disc accretion rates inferred in FUORs are much larger than those in EXORs. 
In the former, typical $\dot M\lesssim 10^{-7}$\MSunPerYear in quiescence, and $\dot M\gtrsim 10^{-5}$\MSunPerYear in outburst. EXORs, on the other hand, only muster $\dot M\sim 10^{-7}$\MSunPerYear during their bursts. Finally, outbursts in FUORs were believed to not repeat on human time scales, whereas EXOR outbursts repeat on time scales of a decade to a few decades \citep[e.g.,][]{Fischer-PPVII,Wang-23-EXLUP,Cruz-23-EXLup}.

A number of authors suggested that the same disc Hydrogen ionisation (thermal instability) physics may work for FUORs \citep{Clarke-90-FUOR,Bell94,BellEtal95,Clarke05-FUORs}. However, FUOR outbursts are observed to last decades to centuries. Disc TI models yield such long outburst duration only if the disc viscosity is $\alpha_{\rm c}\sim 10^{-4}$ on the cold and $\alpha_{\rm h}\sim 10^{-3}$ on the hot branches, respectively. Furthermore, outburst rise times in these models are too long, and the active (that is, participating in the instability) part of the disc is only $R_{\rm act}\lesssim (0.05-0.2)$~au. SED modelling and early FUOR disc interferometry suggested that the active size in observed discs is larger, $R_{\rm act} \sim (0.5-1)$~au \citep{ZhuEtal07,Zhu08-FUOri,Zhu09-FUOri-obs,Eisner11-FUOR}. It has therefore been concluded that the classic TI scenario does not work for FUORs, mainly because their outbursts are too long and their active discs are too large \citep[e.g.,][]{ZhuEtal07}  compared with what theoretical models predict \citep[see also \S 6.1.2 in][]{Armitage15-review}.

However, much more systematic sky monitoring with a wide array of instruments, such as VVV \citep{VVV-Survey-2010}, ASSASN \citep{Shappee13-ASSASN}, ZTF \citep{Masci19-ZTF}, {\em Gaia} \citep{Gaia-2016}, and infrared intereferometry with the MATISSE/VLTI instrument \citep[e.g.,][]{Matisse-telescope-22} provide ample motivation to re-evaluate these conclusions:

\begin{itemize}
    \item Recent interferometry of FU Ori shows that the active disc size is smaller than previously believed, $R\lesssim 0.3$~au \citep{2022Lykou,Bourdarot-23-FUOR}. 
    \item Multi-wavelength observations of the last decade enabled more detailed SED modelling of FUORs than was possible before. Such modelling also points to active disc radii being not too dissimilar from TI predictions. For example, $R_{\rm act} \sim 0.2$~au in HBC722 \citep{Kospal-16-HBC722}, $R_{\rm act} \sim 0.1$~au in Gaia18dvy \citep[Section 4.3 in ][]{Szegedi-Elek-20-Gaia-18}. In fact, \cite{Zhu08-FUOri} concluded $R_{\rm act} \sim 0.25$~au in the classic FUOR V1515 Cyg.
    \item Recent all sky surveys uncovered many unusual eruptive YSOs whose outbursts are just as bright as those of classic FUORs but are surprisingly short, and also show a mixture of spectral characteristics \citep[see][who propose to term such sources ``MNORs"]{Contreras-Pena-17-MNORs}. For example, the bursts in Gaia21bty lasted only half a year \citep{Siwak-23-Gaia-21bty}; Gaia20eae \citep{CruzSaenz-22-Gaia20eae} underwent two outbursts with duration of about a year each, separated by just 7 years. 
    \item Thermal Instability models predict ``reflares": smaller flares with diminishing amplitude on the decaying part of the bursts \citep[e.g.,][]{Coleman16-TI}. We will argue below that such reflares may have been observed in PTF14jg and in Gaia21bty.
\end{itemize} 

Additionally, recent theoretical work also provides motivation to study TI in protoplanetary discs further. \cite{Nayakshin-23-FUOR,Nayakshin23-FUORi-2} showed that if a massive (several Jupiter masses, $\mj$) young planet migrates all the way into the innermost $\sim 0.1$~au, then during a TI outburst the disc is so hot that the planet can start losing its mass rapidly. The planet then becomes the dominant source of matter for the inner disc, powering it for as long as centuries. The planet presence also extends the TI-affected zone outward to $0.3$~au, as observed. This model is intimately related to the scenario proposed by \cite{VB06,VB10} for FUORs but specifically requires TI as a trigger for a quasi steady-state planet mass loss \citep[tidal disruptions of planets also yield accretion outbursts but they are too short and too bright compared with the classic FUORs, e.g.,][]{NayakshinLodato12}. TI is also crucial to explain the fast rise in this scenario.

These observational and theoretical developments motivate our study of the classic TI in protoplanetary discs in this paper. While our numerical methods are inspired by earlier work in the field \citep[such as][]{Bell94,LodatoClarke04}, we use disc viscosity prescription motivated by the observations and modelling of TI in DNe and LMXRBs rather than what is ``needed" to explain the classic FUOR outbursts. In this paper we assume no massive planet is present in the disc. Notwithstanding the scenario of \cite{Nayakshin-23-FUOR,Nayakshin23-FUORi-2} for FU Ori, such planet-free discs should be the norm rather than exception. It is therefore important to ask what the properties of classic TI bursts would be, and how this would compare to the available FUOR/EXOR observations.

The paper is structured as following. In \S \ref{sec:methods} we present our numerical methods. \S \ref{sec:general} presents several model lightcurves, disc properties through an outburst cycle, discusses modelling uncertainties, and  how the salient outburst properties scale with parameters of the problem. In \S \ref{sec:Unusual_fuors} we apply these methods to several young eruptive variable stars with characteristics mixed between FUOR/EXOR classes. Finally, \S \ref{sec:discussion} presents a broad discussion of how these results may relate to the episodic accretion phenomenon in YSOs in general.

\section{Numerical methodology}\label{sec:methods}

\subsection{1D azimuthally integrated equations}

The 1D code we use follows standard approaches to 1D disc modelling of instabilities in dwarf novae and protoplanetary discs \citep[e.g.,][]{Faulkner83-DNe,Cannizzo93-DNe,Bell94,Hameury98-alpha,Zhu10-DZ-MRI-1D,Bae13-MRI-1D}. The code was previously described in \cite{NayakshinLodato12,Nayakshin22-ALMA-CA,Nayakshin-23-FUOR} where we also included disc-planet interactions. Here we omit these terms. The disc surface density $\Sigma$ is evolved in time via
\begin{equation}
    \frac{\partial\Sigma}{\partial t} = \frac{3}{R} \frac{\partial}{\partial R} \left[ R^{1/2} \frac{\partial}{\partial R} \left(R^{1/2}\nu \Sigma\right) \right] 
\label{dSigma_dt}
\end{equation}
where $\nu=\alpha c_{\rm{s}} H$ is kinematic viscosity \citep{Shakura73}, $c_{\rm{s}}$ is the midplane sound speed, and $H = c_{\rm s} \Omega^{-1}$ is the disc vertical scale height; $\Omega = (GM_*/R^3)^{1/2}$ is the Keplerian angular frequency at radius $R$ for stellar mass $M_*$.

We previously used a simplified approach to the thermal balance equation for the disc, following the ideas of \cite{LodatoClarke04}, where two main terms are included: the vertical radiative cooling in the one zone approximation, and the radial advection term. Here we improve this scheme in two ways, following the energy equation form as in \cite{Faulkner83-DNe,Cannizzo93-DNe,Menou99-transition-fronts}, although we recast it in a slightly different form. Instead of the one zone approximation to vertical radiative cooling we used in \cite{Nayakshin-23-FUOR,Nayakshin23-FUORi-2}, here we use  tabulated vertical balance calculations (cf. \S \ref{sec:Scurves}). The disc central (midplane) temperature, $T_{\rm c}$, is evolved according to
\begin{equation}
         \frac{\partial T_{\rm c}}{\partial t} = - \frac{T_{\rm c}-T_{\rm eq}}{t_{\rm cool}} - \frac{1}{R} \frac{\partial}{\partial R}\left( T_{\rm c} R v_r  \right) + 
         \frac{3\nu }{R} \frac{\partial}{\partial R} \left( R \frac{\partial T_{\rm c}}{\partial R}  \right) 
         \;,
    \label{energy_equation}
\end{equation}
where $t_{\rm cool}$ is the local disc thermal (cooling) time scale and $T_{\rm eq}$ is the equilibrium disc temperature, both discussed below. The second term on the right is the  radial heat advection, with $v_r$ equal to the local disc radial velocity. The third term on the right accounts for the radial viscous heat flux. Following \cite{Cannizzo93-DNe} we omit the radial radiative diffusion term, which is of the order of the viscous heat flux; however its addition to eq. \ref{energy_equation} makes it prone to numerical instabilities. \bref{The exact form of the energy equation is however important in the phenomenon of reflares (\S \ref{sec:reflares}), so we plan to explore this issue in future work.}
Furthermore, we found that for the problem parameters explored here, the radial radiative diffusion term typically leads to only $\sim (10-20)$\% changes in the outburst lightcurves. This is much smaller than the lightcurve sensitivity to the viscosity prescription (cf. \S  \ref{sec:example} and \ref{sec:reflares}), which we do vary below.

Equations \ref{dSigma_dt} and \ref{energy_equation} are integrated in an explicit fashion, making sure that time steps are small enough to prevent large changes in $\Sigma$ and other disc variables during one time step. We use a radial grid  uniform in logarithm of $R$  between the inner boundary at $R_{\rm in} = R_*$ and $R_{\rm out}$ set at a few au, with the number of radial bins between  $N_r = 100$ and 200. Following \cite{Bell94} we use the zero torque inner boundary condition. At the outer boundary we use a reflecting boundary condition; this is justifiable since the outbursts never reach $R_{\rm out}$. To achieve quasi-steady state TI cycles we inject mass into the computational domain close to $R_{\rm out}$ at a fixed rate $\dot M_{\rm feed}$.

Physical parameters of a problem are: $M_*$, $R_*$, the disc viscosity prescription discussed below, opacity, and $\dot M_{\rm feed}$. We start our simulations with $\Sigma(R)$ and $T(R)$ obtained by solving the vertically integrated one-zone steady-state disc equations of \cite{Shakura73}. For $M_*\sim 1\, \msun$, and $\dot M_{\rm feed}$ larger than a few $\times 10^{-7}$\MSunPerYear the disc quickly becomes unstable and displays TI cycles.

\subsection{One zone approximation to thermal balance}\label{sec:One_zone}

In the past work we have used the one vertical zone approximation to disc thermal balance, which we describe here for completeness. To solve eq. \ref{energy_equation} we need to specify $t_{\rm cool}$ and $T_{\rm eq}$. The latter is the disc central temperature in  steady state at a given value of $\Sigma$, viscosity parameter $\alpha$, and disc opacity coefficient $\kappa$ (both of which can be functions of $T$ and $\Sigma$), provided all the radial energy flux terms are negligible. In the one zone approximation, the one sided local radiation flux from the disc is
\begin{equation}
    F_{\rm rad} = \sigma_{\rm B} T_{\rm eff}^4 \approx \sigma_{\rm B} 
    \frac{T_{\rm eq}^4}{\tau + \tau^{-1}}\;,
    \label{Frad0}
\end{equation}
where $\tau = \kappa \Sigma/2$. The $\tau^{-1}$ term on the bottom accounts for the situation where the disc is optically thin. In radiative equilibrium,
\begin{equation}
    2 F_{\rm rad} = Q_{\rm visc} = \frac{3}{4\pi} \nu \Sigma \frac{G M_*}{R^3} = 
    \frac{3}{4\pi} \nu \Sigma \Omega \;,
    \label{RadBalance}
\end{equation}
where $Q_{\rm visc}$ is the rate of viscous heating of the disc. As $\kappa$ is a strongly nonlinear function, we solve eqs. \ref{Frad0} and \ref{RadBalance} for $T_{\rm eq}$ iteratively. The cooling time is defined as 
\begin{equation}
    t_{\rm cool} = \text{ min}\left[E_{\rm disc}/(2 F_{\rm rad}), E_{\rm disc}/Q_{\rm visc}\right]\;,
\end{equation}
where $E_{\rm disc}$ is the vertically integrated surface energy density of the disc.

\subsection{Vertical Thermal Balance (S-curves)}\label{sec:Scurves}


The main fallacy of the one zone approach is  the assumption that the opacity coefficient is constant within the disc, equal to its value evaluated at the disc midplane. In fact, $\kappa$ is a strong function of both density and temperature, especially between $T\approx 1,500$~K and a few $\times 10^4$~K. In this temperature range, the disc surface (effective) temperature can be comparable to $T_{\rm c}$ in disc regions of relatively low optical depth, but be an order of magnitude lower than $T_{\rm c}$ in the regions of high optical depth. It is hence essential to take the variation of $\kappa$ with vertical height in the disc into account.

Echoing the modelling approach by a number of previous studies \citep[e.g.,][]{Bell94,Hameury98-alpha,Hirose14-Convection}, in this paper we solve the $z$-dependent vertical balance equations and then organise the results in look-up tables. While more accurate than the one-zone approach described in \S \ref{sec:One_zone}, such calculations are also subject to microphysical, astrophysical and numerical uncertainties. To name a few, the opacity coefficient in the inner disc depends strongly on the dust grain size distribution, composition, molecular and elemental abundances -- all of which are model dependent. For example, the radial drift of grains increases dust abundance in the inner disc \citep[e.g.][]{Vorobyov-Elbakyan-19}, while planetesimal formation may reduce it.  Another unsolved issue is how disc viscous heating and viscosity parameter $\alpha$ depend on height $z$ within the disc \citep[e.g., as noted by][]{Scepi-19-DNe}; so far in the 1D models of TI0unstable discs it has been universally assumed that $\alpha$ is independent of $z$. While 3D radiative transfer MHD calculations can now resolve these issues from first principles \citep{Hirose14-Convection,Hirose15,Scepi-18-TI-alpha}, these calculations themselves take as input global/initial magnetic field configuration in the disc, opacity law, dust abundance, and chemistry at lower temperatures where the gas is partially ionised.

Therefore, pursuing a pragmatic approach in computing the disc vertical thermal balance, we neglect convective energy transfer in the vertical direction. Our calculation follows the approach described in the Appendix of \cite{Hirose14-Convection}. Convection is found to generally transport some few to $\sim 10$\% of energy in 3D calculations except for the region close to the upper unstable point of the S-curve \citep[e.g.,][]{Scepi-18-convection}. Within the uncertainties outlined above, omitting convection is justified as it was shown to yield equilibrium S-curves in good agreement with the 3D radiative MHD calculations  provided that one chooses appropriate values for $\alpha_{\rm cold}$ and $\alpha_{\rm hot}$ \citep[see][]{Hirose15}.

For given values of $M_*$, radius $R$, and disc $T_{\rm eff}$, we start by guessing  the location $z= H_1$ of the $\tau(z)=2/3$ surface where the temperature $T(z) = T_{\rm eff}$. The optical depth $\tau(z)$ is counted from $z=\infty$ downward to the coordinate $z$. The radiative flux of the disc is $F_{\rm rad} = \sigma_{\rm B} T_{\rm eff}^4$. Within the disc,
\begin{equation}
    F(z) = F_{\rm rad} \frac{z}{H_1}\; \quad \text{for } |z| \leq H_1\;.
\end{equation}
The hydrostatic balance equation reads
\begin{equation}
    \frac{dP}{dz} = - \Omega^2 \rho(z) z\;,
\end{equation}
where $P(z)$ and $\rho(z)$ are the local gas pressure and density, respectively. We use an appropriate equation of state \citep[EOS; cf.][]{Nayakshin-23-FUOR} to find $P$ at a given $\rho$ and $T$.
The temperature satisfies the equation
\begin{equation}
    \frac{d T(z)}{dz} = - \frac{3 \kappa(\rho, T) \rho}{16 \sigma_{\rm B} T^3} F(z)\;.
    \label{dTdz_rad}
\end{equation}
We use opacities of \cite{Bell94} by default (we also tabulated S-curves for opacities of \cite{ZhuEtal09} and discuss the differences in \S \ref{sec:reflares}).

We solve these equations downward from $z=H_1$ to $z=0$, iterating to satisfy the vertically integrated energy balance equation,
\begin{equation}
    \frac{Q_{\rm visc}}{2} = \frac{3}{2}\alpha \Omega_k \int_0^{H_1} P(z) dz = F_{\rm rad}\;
    \label{visc_equilibrium}
\end{equation}
for a specific value of $\alpha$. Repeating these calculations for a range of viscosity parameter values from $\alpha = 10^{-4}$ to $\alpha = 10$, we create lookup tables. 

\subsection{Scaling of S-curves with $R$ and $M_*$}\label{sec:scaling}

\begin{figure*}
\includegraphics[width=0.49\textwidth]{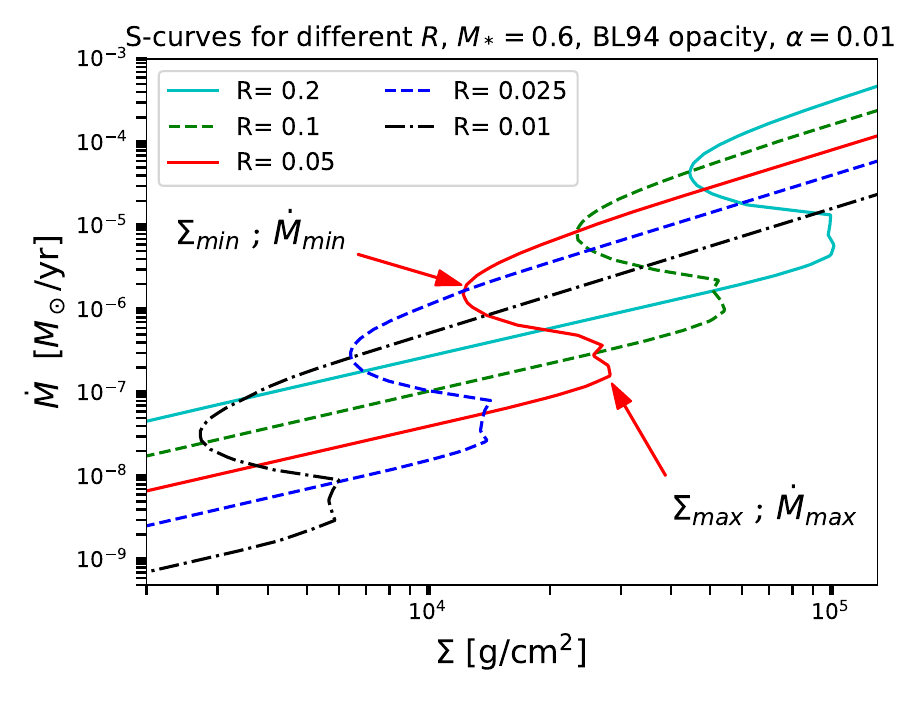}
\includegraphics[width=0.49\textwidth]{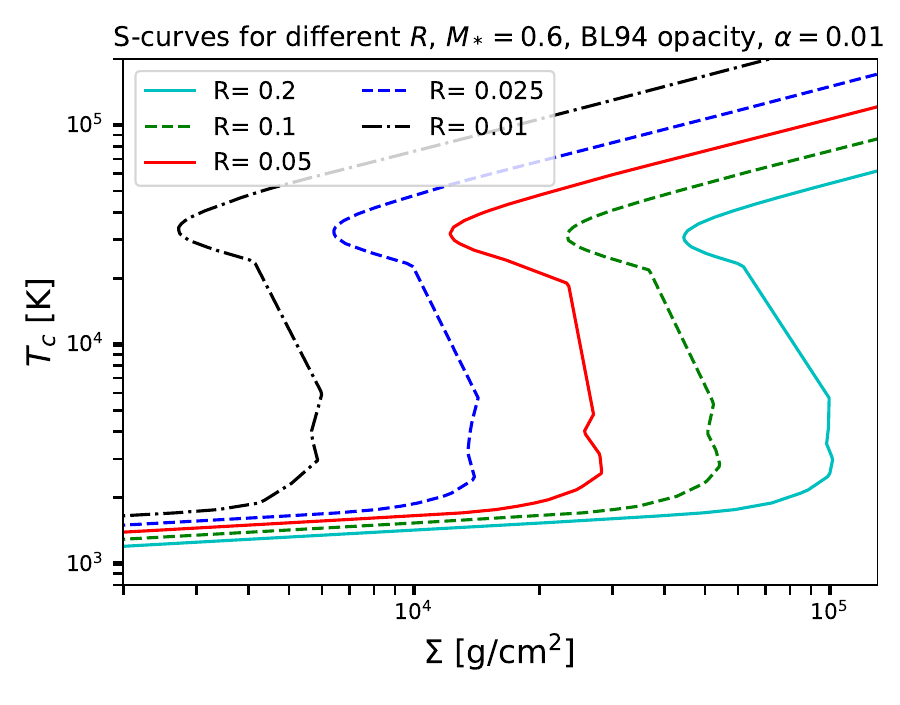}
\caption{Left: Equilibrium $\dot M$ vs disc column depth, $\Sigma$, for the Bell \& Lin (1994) opacities,  mass $M_* =0.6\msun$, fixed $\alpha=0.01$ and various values of radius $R$. Two critical points where the instability sets in are marked with arrows for $R=0.04$~au curve. This disc annulus is unstable to TI if the long-term feeding rate $\dot M_{\rm feed}$ is between $\dot M_{\rm max}$ and $\dot M_{\rm min}$. Right: the same S-curves but with the central (midplane) disc temperature on the vertical axis. See text in \S \ref{sec:scaling} for details.} 
\label{fig:S-Curve}
\end{figure*}

The left panel of Fig. \ref{fig:S-Curve} shows S-curves computed for constant $\alpha=0.01$ and a range of radii from $R=0.01$~au to $R=0.2$ au for a protoplanetary disc around a $M_* = 0.6\msun$ star. The vertical axis shows the steady state accretion rate that is related to the disc cooling flux as (neglecting the boundary condition term [$1-(R_*/R)^{1/2}] \approx 1$ at $R\gg R_*$) 
\begin{equation}
    \sigma_{\rm B} T_{\rm eff}^4 = \frac{3}{8\pi} \frac{G M_* \dot M}{R^3}\;.
    \label{Mdot_ss}
\end{equation}
The right panel of Fig. \ref{fig:S-Curve} shows the same curves but with the disc central (midplane) temperature on the vertical axis. 

To the right of an S-curve, the disc is too cold for a given value of $\Sigma$, i.e., $T <T_{\rm eq}$ since $Q_{\rm visc} > 2 F_{\rm rad}$, so it would heat up to reach steady state. To the left of the S-curve, the inverse is true. The middle section of the S-curve, where $d \dot M/d\Sigma < 1$, is thermally unstable. An annulus of the disc placed on the unstable part of the S-curve could either heat up or cool down if there is a temperature perturbation in the respective direction. In real discs, the two radial energy terms on the right hand side of eq. \ref{energy_equation} decide the direction of travel. These terms are small in a steady state disc compared with the first term in eq. \ref{energy_equation}. However, near the unstable part of the S-curve they matter, making different disc annuli to work together to result in the well known cyclic behaviour \citep[e.g., see Fig. 1 in][]{LodatoClarke04}. 

For an example, consider the S-curve computed for $R=0.05$~au (red solid curve). We marked two important points on this curve with red arrows. For a disc with $\dot M_{\rm feed} < \dot M_{\rm max} \approx 10^{-7}$\MSunPerYear, the disc annulus at this radius is able to find an equilibrium corresponding to a point on the cold stable branch of the S-curve with coordinates $\dot M = \dot M_{\rm feed}$ and $\Sigma = \Sigma(\dot M_{\rm feed})$. Consider now a disc fed at a higher rate, $\dot M_{\rm max} < \dot M_{\rm feed} < \dot M_{\rm min}$. This case is thermally unstable. The cyclic behaviour proceeds in this case as following. Suppose that the annulus first starts at the cold branch of the S-curve. Because $\dot M  < \dot M_{\rm feed}$ there, the disc gains mass from outside faster than it transfers it onto the star, so $\Sigma$ must increase with time. When $\Sigma$ exceeds $\Sigma_{\rm max}$, no local thermal equilibrium is possible, so a transition upward onto the hot stable branch of the S-curve must take place. This transition occurs very rapidly, thus $\Sigma\approx$ constant during it, and this results in an increase of $\dot M$ to about $\sim 10^{-5}$\MSunPerYear in this example. Since $\dot M_{\rm feed}$ is less then this equilibrium $\dot M$ value on the hot branch, the annulus starts to lose mass into the star more rapidly than it gains from larger radii, so $\Sigma$ starts to drop. The local disc annulus now slides down along the hot stable branch of the S-curve towards the second turning point, with coordinates ($\Sigma_{\rm min}$, $\dot M_{\rm min}$). As it reaches that point it still needs to continue the direction of travel, down in both coordinates, but the only stable solution at that point is on the cold stable branch. The disc annulus makes a rapid transition to $\dot M \approx $ a few $\times 10^{-8}$\MSunPerYear in this example. Now $\dot M_{\rm feed} > \dot M$, and the disc starts climbing up the S-curve towards the $\Sigma_{\rm max}$, $\dot M_{\rm max}$ point again.

This single annulus argument would seem to suggest that if $\dot M_{\rm feed} > \dot M_{\rm min}$, then the disc can be stably accreting on the hot branch of the S-curve. However, non-local effects make the disc unstable. Consider, say, $\dot M = 10^{-5}$\MSunPerYear. For the $R=0.05$~au annulus, this corresponds to stable accretion on the hot branch. At the same time, this $\dot M$ places the $R=0.1$ annulus on the unstable part of its local S-curve. Non-local calculations \citep[e.g.,][]{Faulkner83-DNe} show that the whole disc inward of some region that is unstable at a given $\dot M_{\rm feed}$ participates in TI cycles. For $\dot M_*\gtrsim 10^{-5}$\MSunPerYear, as typical of outbursting states of FUORs, radii in the disc up to $R\sim 0.2$~au are unstable.

On the other hand, below a {\em global} critical value for $\dot M_{\rm feed}$, a protoplanetary disc is everywhere stable to TI. As shown by \cite{Bell94}, the instability is usually triggered at $\sim 2 R_*$. If $R_*\sim 3\rsun$, and $\alpha_{\rm c}=0.01$ as in Fig. \ref{fig:S-Curve}, the trigger point corresponds to $R\sim 0.03$~au. Thus at $\dot M_{\rm feed}$ just exceeding a few $10^{-7}$\MSunPerYear the disc is unstable to TI; but the instability region is just a few $R_*$, and non local disc effects can quench the instability from occurring. We will come back to the issue of the critical value for TI triggering in \S \ref{sec:param_space}.

The right panel of Fig. \ref{fig:S-Curve} shows that 
the S-curves look quite similar for all radii in the coordinates $T_{\rm c} - \Sigma$, except for the shift in $\Sigma$. This is because the physics of the S-curve is set by gas opacity and Hydrogen ionisation \citep[e.g.,][]{Hameury-20-review}, and both of these depend on temperature more strongly than on gas density \citep[e.g., cf. the opacity table in][]{Bell94}. As a result,  the point ($\Sigma_{\rm max}$, $\dot M_{\rm max}$) correspond to $T_{\rm c} \approx (2-4)\times 10^3$~K, $T_{\rm eff} \sim 2\times 10^3$~K, whereas the upper turning point on the S-curve, ($\Sigma_{\rm min}$, $\dot M_{\rm min}$),  corresponds to $T_{\rm c} \sim 30,000$~K for all radii. \cite{Hameury-20-review} emphasises that this does not depend on $\alpha$. 

The critical points of the S-curve follow power-law scalings with radius and stellar mass rather well \citep[e.g.,][]{Bell94,Hameury98-alpha,Hameury-20-review}. The fact that the lower turn in the S-curve occurs at $T_{\rm eff} \sim 2,000$~K for typical protoplanetary conditions implies that the quantity $M_* \dot M/R^3$ should be almost invariant at the lower turning point of the S-curve. 
We  find the following power-law scaling for the critical points of the S-curves we computed 
\begin{align}
\Sigma = \Sigma(M_0, R_0) \left( \frac{R}{R_0}\right)^{0.96} \left( \frac{M_*}{M_0}\right)^{-0.3} \alpha_{-2}^{-0.8} \\
\dot M = \dot M(M_0, R_0) \left( \frac{R}{R_0}\right)^{2.425} \left( \frac{M_*}{M_0}\right)^{-0.8} \\
T_{\rm eff} = T_{\rm eff}(M_0, R_0) \left( \frac{R}{R_0}\right)^{-0.14} \left( \frac{M_*}{M_0}\right)^{0.05}
\label{S-curve_scaling}
\end{align}
here $\Sigma(M_0, R_0)$, $\dot M(M_0, R_0)$, $T_{\rm eff}M_0, R_0)$ are the corresponding quantities on the S-curve for stellar mass $M_* = M_0$ and radius $R_0$, $\alpha_{-2} = \alpha_{\rm c}$ divided by $0.01$. For example, for the lower turning point, and for $M_0 =0.6 \msun$ and $R_0=0.1$~au, they are $\Sigma_{\rm max} = 6\times 10^4$ g/cm$^2$, $\dot M_{\rm max} = 10^{-6}$\MSunPerYear, and $T_{\rm eff, max} = 1,350$~K. The power-law exponents in the scaling relations (\ref{S-curve_scaling}) are quite close to those derived by \cite{Hameury98-alpha} and \cite{LodatoClarke04}. 


\section{Properties of classic TI outbursts}\label{sec:general}

The results of this section are not particularly new given previous work of, e.g., \cite{Bell94,Hameury98-alpha,Coleman16-TI} and many others. Our  aims are two fold: (i) review characteristic features of classic TI bursts that could be used to deduce their relevance, or otherwise, to FUOR phenomenon; (ii) highlight the boundaries to which the 1D modelling of classic TI can be considered predictive.

\subsection{Example outbursts}\label{sec:parameters}

In this section we consider some TI examples. We emphasise that the detailed variation of disc viscosity with local disc conditions remains very much model dependent despite all the progress in the field \citep[e.g.,][]{Scepi-19-DNe,Hameury-20-review}. In this paper we follow the \cite{Hameury98-alpha} function to simulate the viscosity parameter increase from the low to the high S-curve branches as a function of disc central temperature:
\begin{equation}
        \ln\alpha(T) = \ln\alpha_{\rm c}\, + \,\frac{\ln\alpha_{\rm h}-\ln\alpha_{\rm c}}{1 + (T/T_{\rm crit})^8}
    \label{alpha_vs_T}
\end{equation}
with $T_{\rm crit}$ is a critical temperature.  For definiteness we pick $M_* = 1.15\msun$ \citep[as deduced for Gaia20eae by][]{CruzSaenz-22-Gaia20eae}, $\dot M_{\rm feed} = 10^{-6}$\MSunPerYear, $\alpha_{\rm c} = 0.01$, $\alpha_{\rm h} = 0.1$, and fix $T_{\rm crit} = 10^4$~K for now.

\begin{figure*}
\includegraphics[width=0.49\textwidth]{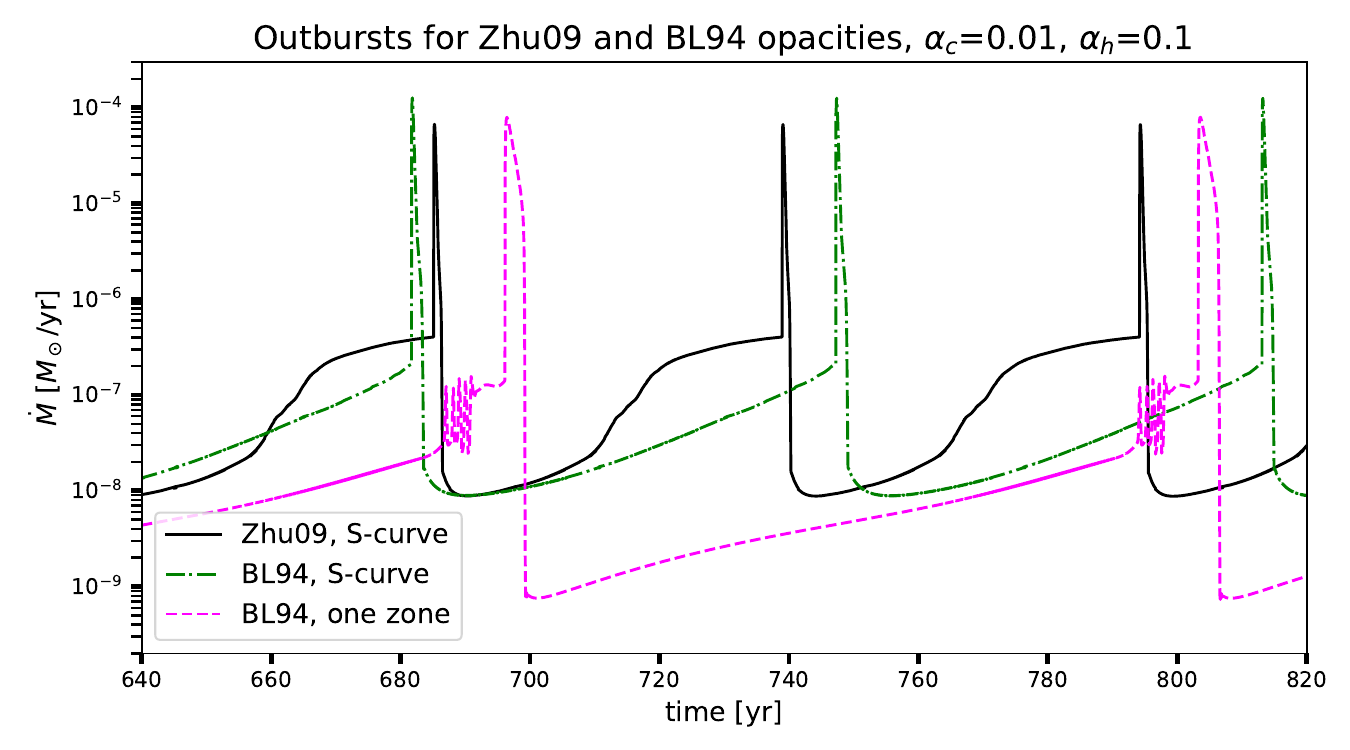}
\includegraphics[width=0.49\textwidth]{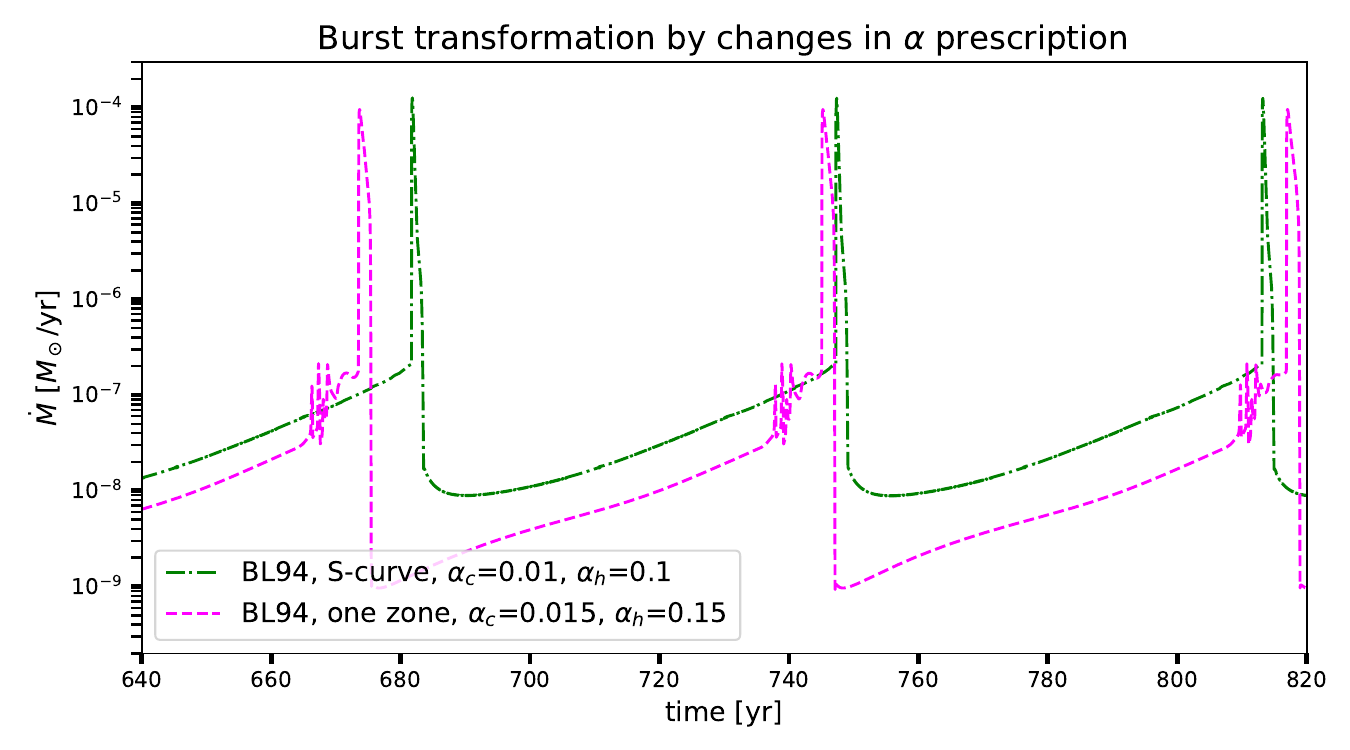}
\caption{Left:  Accretion rate history computed for three different options for the disc thermodynamics, as shown in the legend. Right: The green curve is the same as in the left panel, whereas the one-zone calculation (magenta) is recomputed for slightly different values of $\alpha_{\rm c}$ and $\alpha_{\rm h}$. This shows that one zone calculations result in bursts that are quantitatively different from those computed with the S-curve approach; however the one zone approach may be able to reproduce the S-curve calculations albeit with slightly different viscosity.} 
\label{fig:Comp_Zhu_to_BL94}
\end{figure*}


The left panel in Fig. \ref{fig:Comp_Zhu_to_BL94} shows stellar accretion rates vs time through several TI cycles for three different disc thermodynamics options. For the black and the green curves, the S-curve approach is used, with the \cite{ZhuEtal09} and \cite{Bell94} opacities, respectively. The magenta curve shows the cruder one zone approximation to the disc cooling computed for the \cite{Bell94} opacities. We observe qualitative similarity between the three cases and a various degree of quantitative similarities. For example, for all three cases the maximum $\dot M$ is similar. The two S-curve cases are further similar in the outburst period, the minimum $\dot M$, and the duration of the bright phase (about a year for both). On the other hand, the outbursts for the \cite{ZhuEtal09} opacities show a ``step rise" prior to the main burst (cf. the right panel of Fig. \ref{fig:Comp_Zhu_to_BL94}). Via some admittedly limited experimentation we found that we cannot introduce similar step rises in the outbursts for \cite{Bell94} opacity. This is likely due the fact that the S-curves computed with different opacities are most different around the opacity gap at $T_{\rm c} \sim 2,000-3,000$~K. This corresponds to the disc accretion rate around the ``step" for the values of $M_*$ and $R_*$ chosen here.

The one zone approximation predicts repetition period about twice as long, a much smaller $\dot M$ in the minimum, and outburst duration of about 3 times longer than the S-curve calculations do. This is a significant difference, however we point out that the uncertainty in the values of $\alpha_{\rm c}$ and $\alpha_{\rm h}$ is even more important in practice. In the right panel of Fig. \ref{fig:Comp_Zhu_to_BL94}, the green curve is the same as in the left panel, whereas the black curve is a one zone calculation for the same parameters except that $\alpha_{\rm c}$ and $\alpha_{\rm h}$ increased by 50\% from those in the left panel. The agreement between the one zone and the S-curve approach is now much closer, although still not perfect. 

This shows that results probably depend on the $\alpha$ prescription more sensitively than on the opacity choice. From this point on we shall use the \cite{Bell94} opacity S-curves for definiteness. 

\subsection{Reflares}\label{sec:reflares}

\begin{figure*}
\includegraphics[width=0.49\textwidth]{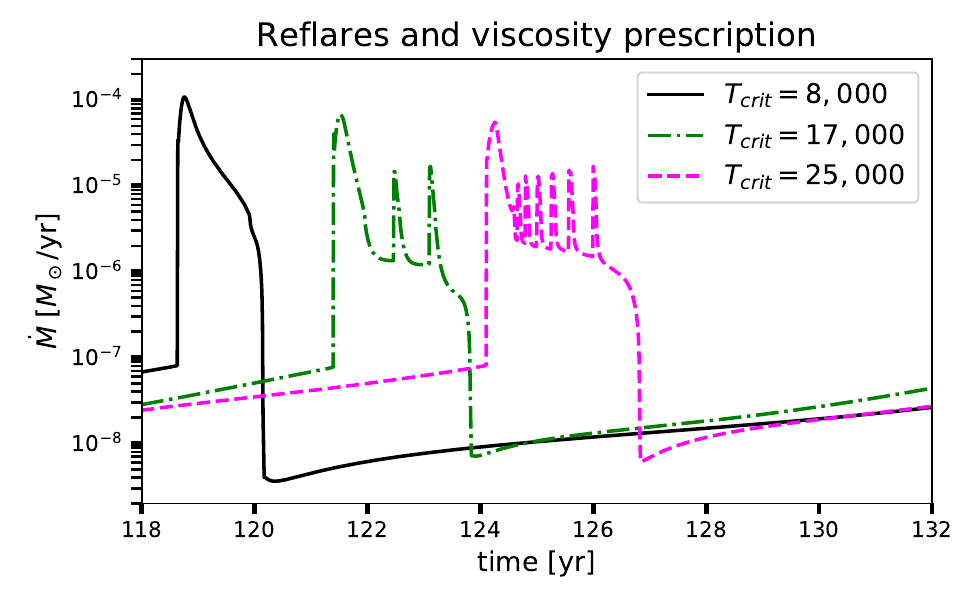}
\includegraphics[width=0.49\textwidth]{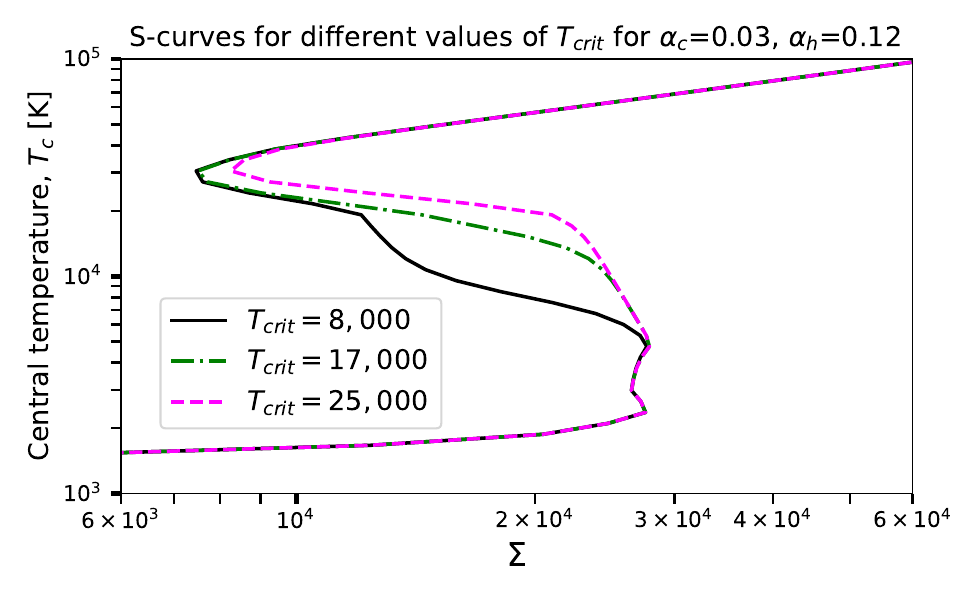}
\caption{Left panel: Accretion rate history for simulations differing only in the parameter $T_{\rm crit}$ in the viscosity prescription (eq. \ref{alpha_vs_T}). The reflares are the secondary peaks appearing after the main outburst peak. Right panel: The S-curves corresponding to the lightcurves shown in the left panel. The higher is the value of $T_{\rm crit}$, the smaller is the difference between the middle unstable and the hot stable branches of the S-curve; this encourages reflare emergence (see text in \S \ref{sec:reflares}).} 
\label{fig:reflares}
\end{figure*}

The left panel of Fig. \ref{fig:reflares} shows accretion rate histories for $M_* = 1.15\, \msun$, $\dot M_{\rm feed} =10^{-6}$\MSunPerYear, $\alpha_{\rm c} = 0.03$ and $\alpha_{\rm h} = 0.15$. The three curves use viscosity prescription given by eq. \ref{alpha_vs_T} but with different values of $T_{\rm crit}$. We see that for the two higher values of $T_{\rm crit}$, there are secondary peaks in the lightcurve after the burst starts to decay from the maximum light. 

Such secondary $\dot M$ peaks are well known in simulations of classic TI and are often called reflares or front reflections. For a thorough review of the subject see \S 4.3 in \cite{Lasota01-Review}, and \cite{Coleman16-TI}. The reflares appear during the inward cooling front propagation after the outburst peak \citep{Menou-00-reflares}. Material flowing in from the outer regions may create a density spike in $\Sigma$ behind the cooling front, and this can re-ignite a smaller secondary outburst in the region that only recently cooled down to the cold branch of the S-curve \citep[see Fig. 13 in][]{Coleman16-TI}. For the experiments in Fig. \ref{fig:reflares}, the secondary trigger point is at $R\sim 0.03$ au. These reflares are a physical not numerical property of the disc models yet they do not seem to be observed in DNe and low mass X-ray binary transients LMXRBTs.

Reflare emergence and observational appearance depends on (1) disc viscosity \citep[][]{Mineshige-85-DNe}; (2) the extent of the unstable region and (3) disc irradiation by the central object \citep{Menou-00-reflares}; (4) outer boundary conditions; and (5) magnetic torques from the central object \citep{Scepi-19-DNe}; and of course (6) the exact shape of the S-curve. For the problem at hand, we do not consider magnetic torques from the star since the inner disc in FUORs apparently proceeds all the way to the star during the bursts
, crushing the magnetosphere \citep[e.g.,][]{HK96}. The irradiation by the central star is also unlikely to affect reflares as the disc heating is dominated by local viscous energy liberation in the protoplanetary disc case in the region where the reflares would occur. The outer disc radius changes, occurring during outbursts in compact binary systems \citep{Menou-00-reflares}, are also not important as FUOR discs are much larger, extending far beyond the TI-unstable disc region.


Reflares were not yet discussed in the context of protoplanetary systems.  \cite{Bell94} \& \cite{LodatoClarke04} did not report reflares in their models. It is possible that this is because they used viscosity parameters smaller by two orders of magnitude than we do; when we assume similarly low values of $\alpha$ the reflares tend to disappear for a wide range of viscosity laws. However, we find that for modern observation and simulation motivated viscosity parameter values \citep[e.g.,][]{King07-alpha,Hirose15,Scepi-18-TI-alpha} reflares are often present as Fig. \ref{fig:reflares} exemplifies.



The right panel of Fig. \ref{fig:reflares} shows the equilibrium S-curves for the three numerical expriments presented in the left panel of the figure.  We see that while the critical points of the S-curve do not appear to move significantly between the three cases, the middle unstable branch of the S-curve does shift significantly towards the upper stable branch when $T_{\rm crit}$ increases. This is apparently important for the existence of reflares. Despite the middle branch being unstable, disc anulli do travel along it temporarily during instability cycles \citep{Smak88-DNe,Bell94,Menou99-transition-fronts}. It is likely that the smaller separation between the stable and the unstable branches for the two higher values of $T_{\rm crit}$ in Fig. \ref{fig:reflares} encourages the emergence of reflares \citep[see \S 4.2 in][for a further discussion]{Coleman16-TI}.

For the rest of the paper we set $T_{\rm crit} = 8,000$~K as a default as this value appears to avoid reflares for a significant range in the values of $\alpha_{\rm c}$ and $\alpha_{\rm h}$. However we shall come back to the issue of reflares in \S \ref{sec:PTF14jg} when discussing PTF14jg.

\subsection{Disc evolution during an outburst}\label{sec:example}

\begin{figure*}
\includegraphics[width=\textwidth]{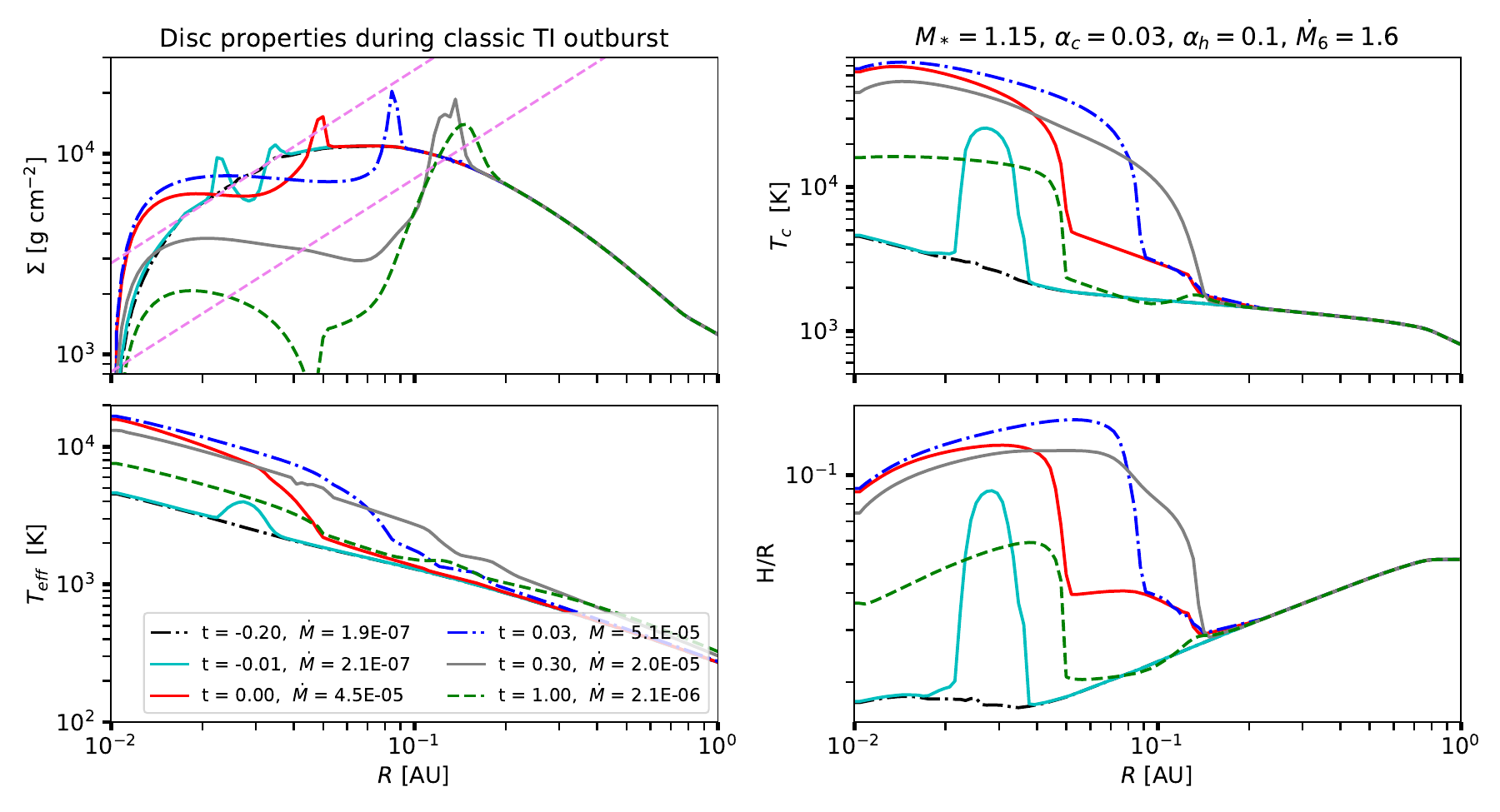}
\caption{Snapshots of the inner disc just before and during one burst of the classic TI corresponding to the blue curve in the middle panel of Fig. \ref{fig:Lots_of_bursts}. The purple dashed lines in the top left panel show the critical disc surface densities (cf. the left panel of Fig. \ref{fig:S-Curve} and eq. 11) on the S-curve as a function of radius. The upper curve corresponds to $\Sigma_{\rm max}$ and the lower one to $\Sigma_{\rm min}$. Above $\Sigma_{\rm max}$ the TI is triggered and the disc travels to the hot (ionised H) stable branch of the S-curve, whereas below $\Sigma_{\rm min}$ the disc falls back to the cold neutral H branch. Note that the instability trigger occurs close to the star and is very rapid. The maximum extent of the TI-affected region is $R_{\rm TI} \sim 0.15$~au. The legend shows the respective times for each snapshot.} 
\label{fig:Classic_TI_Sigma_T}
\end{figure*}

Fig. \ref{fig:Classic_TI_Sigma_T} shows an example of how accretion disc properties change during one outburst. These results echo similar conclusions by \cite{Bell94} albeit the significant difference in time scales brought upon by the much larger values of $\alpha$ we use. This section will be useful for us in \S \ref{sec:param_space}, and for the general discussion of TI relevance to FUORs later on. The outburst we picked for this analysis is the one shown with the blue dashed curve in the bottom panel of Fig. \ref{fig:Lots_of_bursts}, with the time axis shifted to have $t=0$ at the moment when the burst trigger first occurs. We define the latter as the moment in time when the accretion rate onto the star rises one-two orders of magnitude from its pre-burst value; this occurs later than the trigger region managing to reach the hot branch of the S-curve.

The first snapshot in Fig. \ref{fig:Classic_TI_Sigma_T} is at $t=-0.2$ year. The legend also shows the corresponding $\dot M$ on the star, which is $\approx 2\times 10^{-7}$\MSunPerYear just before the burst. The $\Sigma$ panel shows two power laws with dashed magenta lines. These show $\Sigma_{\rm max}$ and $\Sigma_{\rm min}$, the two critical turning points on the S-curve (recall that these scale with $R$ as per eqs. \ref{S-curve_scaling}). Before the trigger occurs, the disc evolves slowly, changing on the viscous time of the cold branch (tens of years). During that time $\Sigma$ creeps up towards the $\Sigma_{\rm max}$ line, and eventually crosses it at $R_{\rm trig} \sim 0.028$~au (2.8 $R_*$). The trigger is obvious at the next time moment in all four panels in Fig. \ref{fig:Classic_TI_Sigma_T}, but especially in the $T_{\rm c}$ and $H/R$ profiles. After the trigger the ionisation fronts propagate both inward and outward. The inward propagating front reaches the star in about a week, in accord with the expected front propagation time \citep[e.g.,][]{Meyer84-transition-fronts,Menou99-transition-fronts}, $R_{\rm trig}/(\alpha_{\rm h} c_{\rm h})$, some $\sim 10$ orbits at $R_{\rm trig}$. Here $c_{\rm h}$ is the sound speed at the disc midplane on the hot branch of the S-curve. The outburst is not yet at the maximum light at that point, however: the next snapshot (blue curves) in Fig. \ref{fig:Classic_TI_Sigma_T} has higher $\Sigma$ and $T_{\rm eff}$ in the innermost disc. 

The burst rise to the maximum is powered by the outer ionisation front moving to larger $R$ which provides new mass supply to the inner disc. Since accretion of gas on the star requires transfer of angular momentum outward, the ionisation front also takes some gas with it -- this is seen in the spike in $\Sigma$ propagating outward. This spike allows disc annuli that were stable before the burst (their $\Sigma$ was lower than $\Sigma_{\rm max}$) to cross the instability threshold. However, as $\Sigma_{\rm max}$ is a strong function of radius, eventually the ionisation front stalls. The maximum extent of the disc which participates in the burst is seen to be close to where the $\Sigma_{\rm min}$ line crosses the pre-burst $\Sigma(R)$ curve, i.e., $R\approx 0.15$~au in Fig. \ref{fig:Classic_TI_Sigma_T}. It is interesting to note that 3D radiation magnetohydrodynamics simulations of \cite{Zhu20-FUOR} -- who do not assume an $\alpha$ viscosity prescription but calculate it from first principles -- also yield the size of the disc region where Hydrogen is ionised to be about 0.2 au (e.g., see their Figs. 6 and 14).

The fall of the burst from maximum begins when the outward moving ionisation front stalls. A cooling front then propagates inward with speed $\sim \alpha_{\rm h} c_{\rm h} (H/R)^{q_c}$, with $q_c \sim 0.5$ to 0.7 \citep[see][]{Vishniac96-cooling-fronts,Lasota01-Review}. Since $H/R \sim 0.1-0.2$ on the hot branch, this indicates that the decaying phase of the burst will be $\sim 5$ times longer than the rise to the maximum light. This is born out in our numerical models, with the rise time generally being of order 0.2 months and burst duration of order a year (both of these time scales do depend on $M_*$, $\dot M_{\rm feed}$, and $\alpha$-prescription). During the cooling front propagation phase the inner disc has a quasi-steady state \cite{Shakura73} structure with a gradually decreasing $\dot M$ \citep{Menou99-transition-fronts}.

\subsection{TI outburst scaling with $\alpha$, $M_*$, $\dot M_{\rm feed}$}\label{sec:param_space}

\begin{figure}
\includegraphics[width=1\columnwidth]{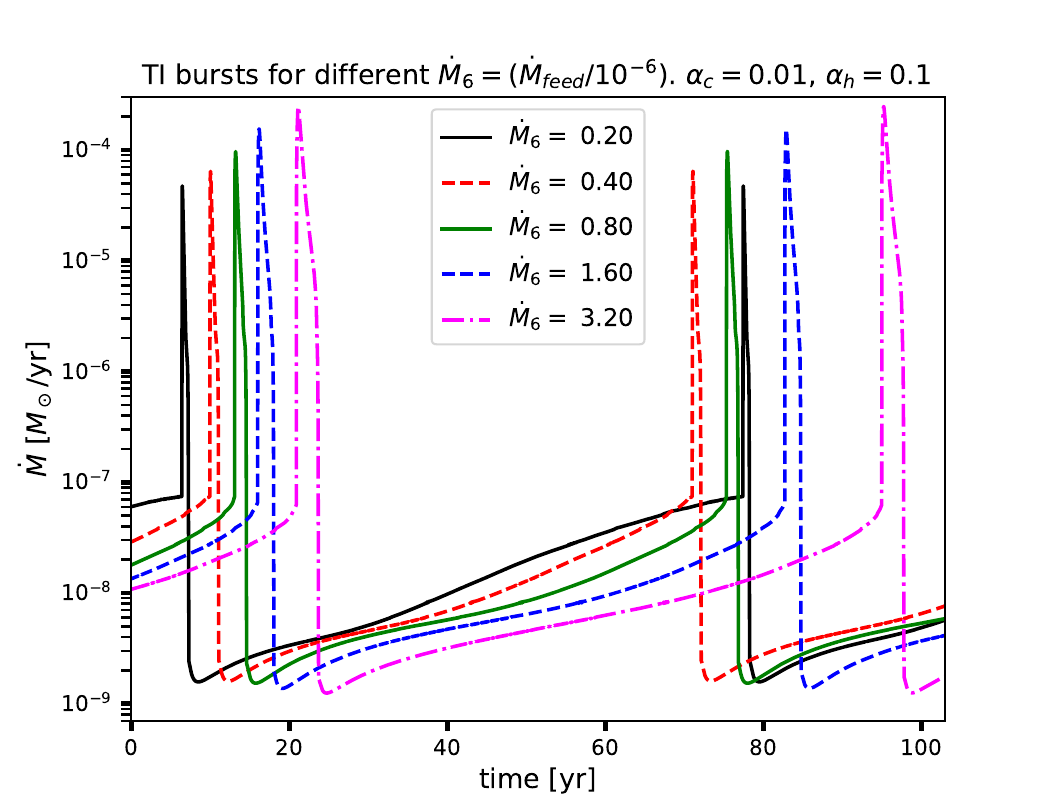}
\includegraphics[width=1\columnwidth]{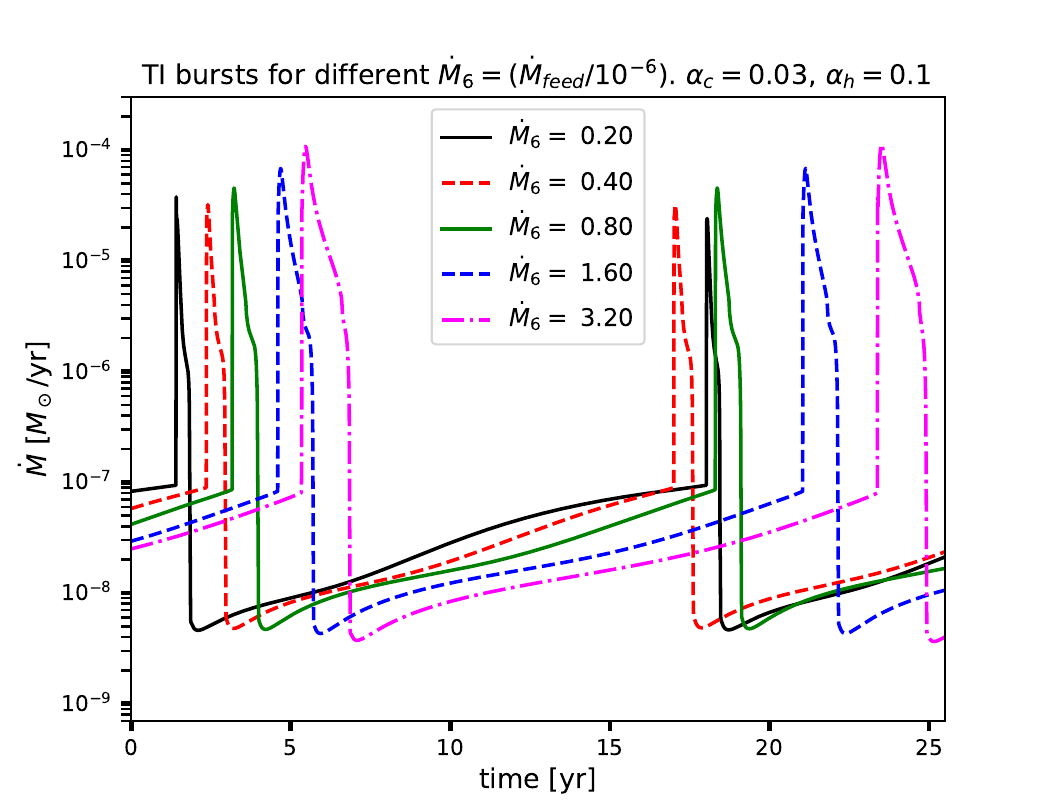}
\includegraphics[width=1\columnwidth]{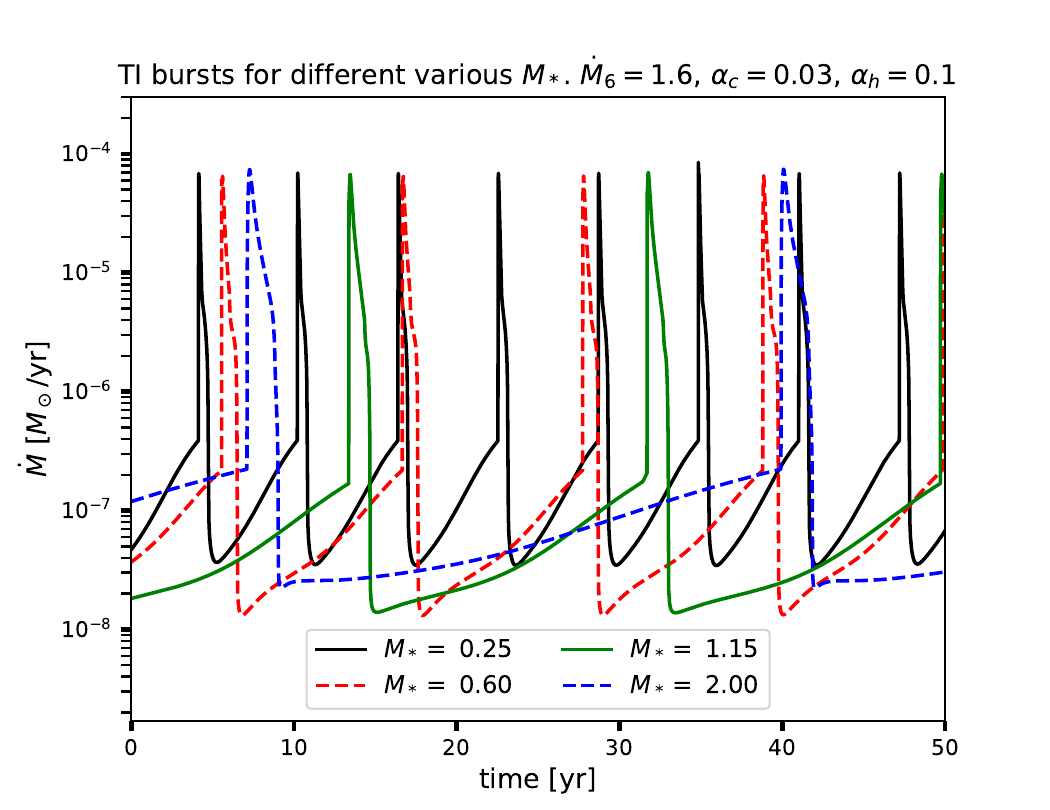}
\caption{Top: Stellar accretion rate history for five different feeding rates and $\alpha_{\rm c} =0.01$, $M_* = 1.15\, \msun$. Middle: Same but for $\alpha_{\rm c} = 0.03$. Bottom: Similar to the above but for a fixed $\dot M_{\rm feed} = 1.6\times 10^{-6}$\MSunPerYear, $\alpha_{\rm c} = 0.03$, and several values of $M_*$. Please note that the time axis is different for the three panels. For further detail and discussion see \S \ref{sec:param_space}.}
\label{fig:Lots_of_bursts}
\end{figure}

Fig. \ref{fig:Lots_of_bursts} shows the accretion rate history for a number of simulations for  $R_* = 0.01$ au ($\approx 2\rsun$), $\alpha_{\rm h} = 0.1$, and a selection of other parameters, such as $M_*$, $\dot M_{\rm feed}$, and $\alpha_{\rm c}$. In the top and middle panels, $M_*=1.15\,\msun$, but the value of $\alpha_{\rm c}$ differs: $\alpha_{\rm c} = 0.01$ and $0.03$, respectively. The disc feeding rate from $\dot M_{\rm feed}$ is varied from $2\times 10^{-7}$\MSunPerYear to $3.2\times 10^{-6}$\MSunPerYear, as shown in the legend. Below $\dot M_{\rm feed} \sim 10^{-7}$\MSunPerYear the thermal instability does not develop for these system parameters as the whole of the disc is on the cold stable branch, so we do not present such low $\dot M_{\rm feed}$ models. 
The bottom panel of Fig. \ref{fig:Lots_of_bursts} shows how burst properties vary with $M_*$ at fixed $\dot M_{\rm feed} = 1.6\times 10^{-6}$\MSunPerYear, and $\alpha_{\rm c}=0.03$. All of these models were computed for $T_{\rm crit} = 8,000$~K and we see that none of them show reflares.

Several trends in the burst properties stand out (please note that the time axis is different for the three panels in Fig. \ref{fig:Lots_of_bursts}):

\begin{enumerate}
    \item Stellar accretion rate in quiescence is always smaller than $\sim$~a few $\times 10^{-7}$\MSunPerYear for system parameters explored in the Figure.
    \item The  repetition time between the bursts, $t_{\rm rep}$, only weakly depends on the feeding rate $\dot M_{\rm feed}$, increasing slowly but not always monotonically with it.
    \item $t_{\rm rep}$ is a strong function of the disc viscosity on the cold branch, roughly scaling with $\alpha_{\rm c}^{-1}$.
    \item $t_{\rm rep}$ strongly anti-correlates with stellar mass; the more massive the star is, the less frequent the bursts are. 
    \item The peak accretion rate on the star, $\dot M_{\rm peak}$, and the burst duration, $t_{\rm b}$, correlate positively with $\dot M_{\rm feed}$.
    \item The bursts are longer for more massive stars.
    \item The minimum accretion rate onto the star are all comparable, decreasing very slightly with $\dot M_{\rm feed}$. This surprising scaling is due to the disc being emptied right after a burst out to a larger radius at higher $\dot M_{\rm feed}$.
\end{enumerate}

These scalings can be understood semi-analytically. \cite{Bell94} show that the size of the disc region undergoing TI is limited to the radius within which the disc effective temperature $T_{\rm eff}$ in the steady-state \cite{Shakura73} solution exceeds $\sim 2000$~K:
\begin{equation}
    R_{\rm TI} \approx 0.065 \text{ au } \dot M_{-6}^{1/3} M_1^{1/3}
    \;,
    \label{R_TI}
\end{equation}
where $\dot M_{6} = \dot M_{\rm feed}/(10^{-6}$ \MSunPerYear), $M_1 = M_*/\msun$. This equation does not depend on $\alpha_{\rm c}$ or $\alpha_{\rm h}$. Crucially, the classic TI instability region shrinks with decreasing $\dot M_{\rm feed}$, so \cite{Bell94} define the critical feeding rate below which TI instability does not occur -- the disc is always and everywhere on the cold stable branch of the S-curve. Here we reproduce their result,
\begin{equation}
    \dot M_{\rm crit} \approx 5\times 10^{-7} \text{ \MSunPerYear} \; \left( \frac{R_*}{3 \rsun}\right)^3\, M_1^{-1}\;,
    \label{dotM_crit}
\end{equation}
including $\dot M_{\rm crit}$ scalings with $M_*$ and $R_*$. If mass-radius relation of central stars of episodic accretors was well known, we could eliminate $R_*$ from this equation. In practice this is complicated by the unknown and probably strongly varying accretion history of the stars, since accretion of gas at high rates can lead to $R_*$ swelling \citep{StahlerEtAl80,Hartmann97-Birth-line,Hosokawa11-L-spread,BaraffeEtal12}. Nevertheless, it is interesting to note that $R_*$ is usually found to increase with $M_*$. For example, based on Fig. 5 in \cite{BaraffeEtal12},  for the accretion shock energy injection efficiency  into the star $\alpha = 0.2$ (see eq. 1 in their paper, not to be confused with the viscosity parameter we use here), $R_*\propto M_*^{\beta_m}$, with $\beta_m\approx 0.43$ in the range $0.26\msun \leq M_* \leq 0.9\msun$. For this scaling we obtain from eq. \ref{dotM_crit} 
\begin{equation}
    \dot M_{\rm crit} \propto M_1^{0.29}\;,
    \label{dotM_crit2}
\end{equation}
which shows a rather weak dependence on $M_*$. This is therefore an important although only approximate prediction of the theory (e.g., the pre-factor in eq. \ref{dotM_crit} depends on the opacity in the disc somewhat, since the S-curve depends on it): episodic accretors with $\dot M \lesssim \dot M_{\rm crit} \sim 5\times 10^{-5}$\MSunPerYear are on the low stable branch of the S-curve. Furthermore, there should be a significant gap between this type of sources and those that did exceed $\dot M_{\rm crit}$ -- these sources then run away to the hot stable branch of the S-curve where $\dot M$ is $\sim 2$ orders of magnitude higher (cf. $\dot M_{\rm peak}$ below). This statistical prediction of classic TI can in principle be tested (cf. \S \ref{sec:discussion_burst_desrt}).

\cite{Bell94} point out that the ionisation front does overshoot radius $R_{\rm TI}$ somewhat (see also \S \ref{sec:example}, Fig. \ref{fig:Classic_TI_Sigma_T}). Nevertheless, eq. \ref{R_TI} can be used to evaluate properties of classic TI bursts approximately. For example, the burst rise time is
\begin{equation}
    t_{\rm rise} \sim \frac{R_{\rm TI}}{\alpha_{\rm h} c_s} \approx 0.16 \hbox{ year } \dot M_{-6}^{1/3} M_1^{1/3} \left(\frac{\alpha_{\rm h}}{0.1}\right)^{-1}\;,
    \label{t_rise}
\end{equation}
while the decay from the maximum time -- and therefore the approximate duration of the burst phase -- is about a factor of $5-10$ longer as noted in \S \ref{sec:example}. The mass within the region unstable to TI is $\sim \pi \Sigma_{\rm max} R_{\rm TI}^2$, 
\begin{equation}
    M_{\rm TI} = 5\times 10^{-5}\, \msun \, \alpha_{-2}^{-0.8} \dot M_{-6}^{0.97} M_1^{0.69} \;.
    \label{M_TI}
\end{equation}
This is the mass available to power classic TI bursts. Assuming that the unstable disc is completely emptied out during the burst and then needs to be refilled at the mass supply rate $\dot M_{\rm feed}$, the repetition time scale is
\begin{equation}
    t_{\rm rep} \approx 51 \;\text{ years } \alpha_{-2}^{-0.8} M_1^{0.7}\;,
    \label{t_rep}
\end{equation}
where we rounded off to the second significant digit, which cancelled out a very weak dependence on $\dot M$. This equation is very much consistent with the scalings with $\alpha$, $M_*$ and $\dot M$ that we see in Fig. \ref{fig:Lots_of_bursts}.

\cite{Bell94}, their \S 6.1.3, show that radial transfer of mass places a strong non-local limit on the maximum accretion rate achievable during the outburst. We can therefore estimate the peak burst accretion rate as the disc mass at the radius $R_{\rm trig}$ divided by the viscous time on the hot branch at that radius:
\begin{multline}
    \dot M_{\rm peak} = 4\times 10^{-5}\; \text{\MSunPerYear } \left(\frac{R_*}{2 \rsun}\right)^{1.5} \alpha_{-2}^{-0.8} M_1^{0.2} \\
    \times \left(\frac{R_{\rm trig}}{2.5 R_*}\right)^{1.5} \left(\frac{\alpha_{\rm h}}{10 \alpha_{\rm c}}\right)
    \left(\frac{H/R}{0.15}\right)^2
    \label{dotM_peak}
\end{multline}
The latter equation assumes that the burst is scaled on $R_{\rm trig} \sim 5 \rsun$, and assumes $H/R \sim 0.15$. These values do evolve a little with $\dot M_{\rm feed}$ in simulations\footnote{Note also that the trigger point scales with $R_*$, which we assumed fixed here.}, and this explains why eq. \ref{dotM_peak} does not quite reproduce the relatively weak correlation between $\dot M_{\rm peak}$ and $\dot M_{\rm feed}$ that we see in Fig. \ref{fig:Lots_of_bursts}. Eq. \ref{dotM_peak} shows that the peak accretion rate does depend on the ratio of $\alpha_{\rm h}$ to $\alpha_{\rm c}$ that we see in our models.

One other quantity of interest is the bolometric luminosity of the disc at the peak of the burst, $L_{\rm peak} = G M_* \dot M_{\rm peak}/R_*$,
\begin{multline}
        L_{\rm peak} = 640\; L_\odot
        \left(\frac{R_*}{2 \rsun}\right)^{0.5} \alpha_{-2}^{-0.8} M_1^{1.2} \\        
    \times \left(\frac{R_{\rm trig}}{2.5 R_*}\right)^{1.5} \left(\frac{\alpha_{\rm h}}{10 \alpha_{\rm c}}\right)
    \left(\frac{H/R}{0.15}\right)^2
    \label{L_peak}
\end{multline}
This equation scales with $M_*$ relatively weakly, which implies that, compared to the stellar photospheric luminosity, the bursts are very bright in Solar type stars, but become undetectable for very high mass stars, $M_*\gtrsim 15\msun$, assuming that stellar luminosity scales with $M_*$ somewhat as it does on the ZAMS, $L_*\sim \lsun (M_*/\msun)^3$ or so. Indeed the maximum luminosity for a star of $20\msun$ given by eq. \ref{L_peak} is only a little larger than $10^4\lsun$.


\section{Thermal Instability in FUORs}\label{sec:Unusual_fuors}

In this section we discuss four recently discovered peculiar FUOR/EXOR bursters. These sources showed visual magnitude increases and/or outburst spectra commensurate with FUORs yet they decayed from the peak light on timescales of a year or so, which is more typical of EXORs. Further, two of the sources show evidence for burst repetition on timescales of order a decade, and there are also hints of a reflare-like behaviour while the sources dim from the peak.

\subsection{PTF14jg}\label{sec:PTF14jg}

\begin{figure}
\includegraphics[width=1\columnwidth]{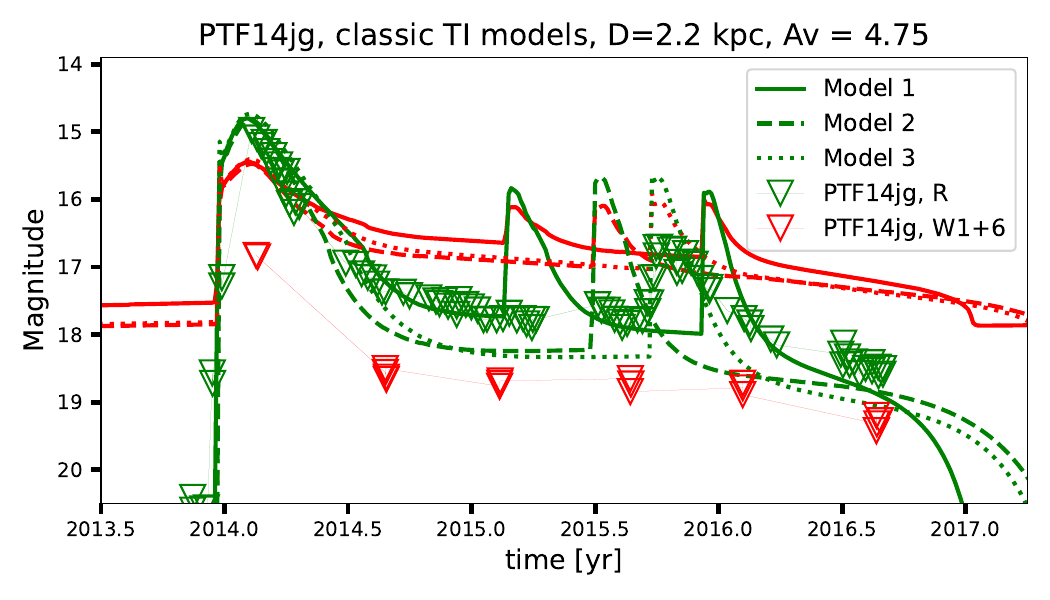}
\caption{The R and W1 band data (effective wavelength $0.651 \mu$m and $3.35 \mu$m, respectively) for the outburst of PTF14jg \citep[triangles; ][]{Hillenbrand19-PTF14} compared to the simulated lightcurves for several classic TI models in the same bands. While our model discs match the R band data reasonably well, they are always too bright in the W1 band compared to the observation.}
\label{fig:PTF14_D2}
\end{figure}

Fig. \ref{fig:PTF14_D2} shows the optical (R band, green triangles) and IR (W1 {\em NEOWISE} band, red stars) lightcurve of PTF14jg \citep{Hillenbrand19-PTF14}, compared to several classic TI models that we discuss in detail below. Prior to its discovery by the Palomar Transient Factory survey \citep{Law09-PTF} in 2013 the source was a faint and virtually uncatalogued star. There are no pre-outburst spectra, and the distance to it is poorly constrained, $D = (2-6)$~kpc, with some preference for the lower value \citep{Hillenbrand19-PTF14}. Likewise, extinction on the line of sight is estimated to be $A_{\rm V} = 4.75$, but lower values are possible. 

As discussed by \cite{Hillenbrand19-PTF14}, PTF14jg has several characteristics of the classical FUORs: (i) the outburst rose by $\sim 6$ magnitudes in the optical over a timescale of a few months; (ii) spectral line features ubiquitous in strong outflows; (iii) low gravity absorption line spectrum systematically changing with spectral type, indicating that the spectrum is probably dominated by emission from a rotating inner disc. At the same time, the outburst lost half of its magnitude increase in less than a year. The outburst duration is much shorter than that of, say, FU Ori  \citep{Clarke05-FUORs,2022Lykou}. PTF14jg also shows evidence for a secondary peak in the second quarter of 2015 that was probably missed when the source was behind the Sun, and a tertiary maximum clearly seen in the last quarter of 2015. Such rapid falls from the maximum light and somewhat chaotic variability is more characteristic of EXOR bursts \citep[e.g.,][]{Wang-23-EXLUP}  than of FUORs.

In our first set of numerical models, we used \cite{Hillenbrand19-PTF14} ``best" values for the distance and extinction to the source, $D\sim 2$ kpc and $A_{\rm V} = 4.75$, respectively. We set a relatively low stellar mass, $M_* = 0.5 \, \msun$, and $R_* = 2\rsun$. We started by aiming to reproduce the R band lightcurve of PTF14jg only. When computing model lightcurves we integrate time-dependent multi-colour blackbody spectrum over the disc surface. Note that $T_{\rm eff}(R)$ curves are not the usually assumed power-laws, they are a result of the time-dependent calculation (e.g., the lower left panel of Fig. \ref{fig:Classic_TI_Sigma_T}).

In this modelling we used \cite{Bell94} opacity and considered the following free parameters: three parameters that describe the S-curve shape, $\alpha_{\rm c}$, $\alpha_{\rm h}$, $T_{\rm crit}$, and the feeding accretion rate $\dot M_{\rm feed}$ being the fourth parameter. We eye-balled the differences between the model R band lightcurves and the observed data. As this is a relatively time-consuming procedure we explored only about two dozen parameter combinations that we adjusted by hand after each try, and also used the outburst scalings discussed in \S \ref{sec:scaling} to guide us in parameter adjustment procedure. Once we found a reasonably close match between the R band model data and the observation we also considered the IR band W1.

\begin{figure}
\includegraphics[width=1\columnwidth]{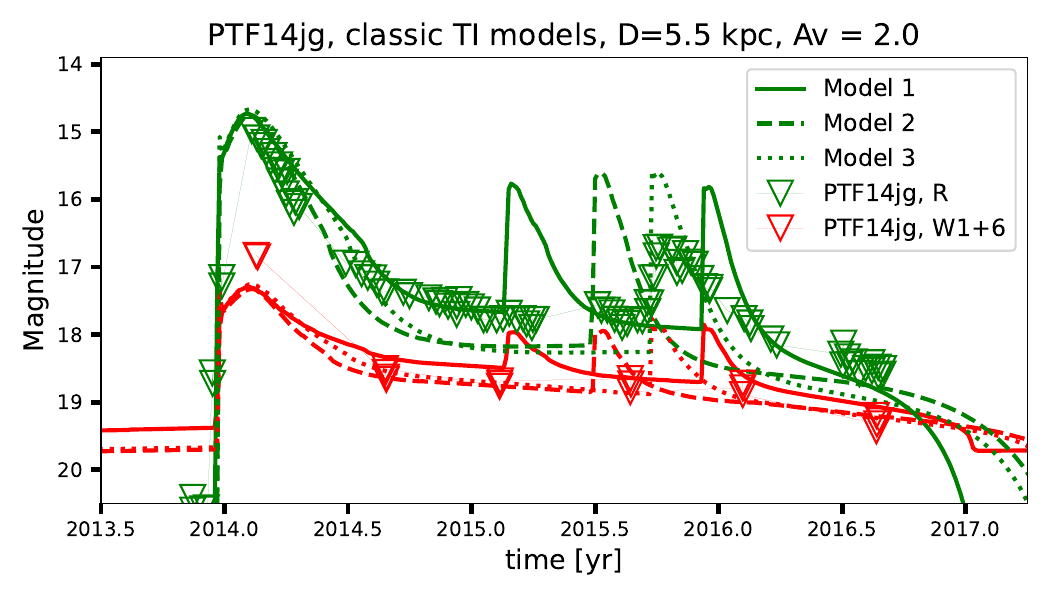}
\caption{Same as Fig. \ref{fig:PTF14_D2} but assuming larger distance and lower extinction. In this case there is a better agreement between the models and observations in W1 band.}
\label{fig:PTF14_D5}
\end{figure}

\begin{table}
    \centering
    \begin{tabular}{c|cccl}
         Model&  $\alpha_{\rm c}$&  $\alpha_{\rm h}$&  $T_{\rm c}$ [K]& $\dot M_{\rm feed}$\\
         \hline
         1&  0.02&  0.04&  $1.5\times 10^4$& $2.0\times 10^{-6}$\\
         2& 0.015 &  0.045 & $1.5\times 10^4$ & $1.5\times 10^{-6}$ \\
         3&  0.015 & 0.045  & $1.75\times 10^4$ & $1.5\times 10^{-6}$\\
    \end{tabular}
    \caption{Models shown in Figs. \ref{fig:PTF14_D2} and \ref{fig:PTF14_D5}.}
    \label{tab:Table1}
\end{table}

Our initial goal was to enquire whether there is a combination of parameters for a classic TI outburst in the R band similar to PTF14jg in the bulk part: a rapid rise, followed by slower but still quite a rapid decline by $\sim$ 3 magnitudes, a plateau, and then weaker secondary flares. The models shown in Fig. \ref{fig:PTF14_D2} reach this goal to a degree; their parameters are shown in Table \ref{tab:Table1}. While our model discs are encouragingly similar to the R band data of PTF14jg, they are $\sim 2$ magnitudes too bright in the IR to fit the data. One explanation for this could be that PTF14jg active disc is surprisingly small. \cite{Liu22-FUORs} find that to account for the absence of absorption lines in the IR spectra of PTF14jg, its active disc must end as close as 0.06 au from the star. We explored this \citep[unphysical, as pointed out by][]{Liu22-FUORs} scenario, but at least for the relatively small number of models explored here this did not remedy the disagreement.

We plan to explore the parameter space of theoretical models more fully in future work. For now we considered another possibility: that extinction at the source is lower than the value $A_{\rm V} = 4.75$ preferred by \cite{Hillenbrand19-PTF14}, and that therefore the distance $D$ is larger. We found that using the same models shown in Fig. \ref{fig:PTF14_D2} but with $D=5.5$~kpc \citep[which is within the range considered by][]{Hillenbrand19-PTF14} and $A_{\rm V} = 2.0$ resulted in a more reasonable agreement between the models and the data in the IR as well.

\subsection{Gaia20eae}\label{sec:Gaia20eae}

\begin{figure}
\includegraphics[width=1\columnwidth]{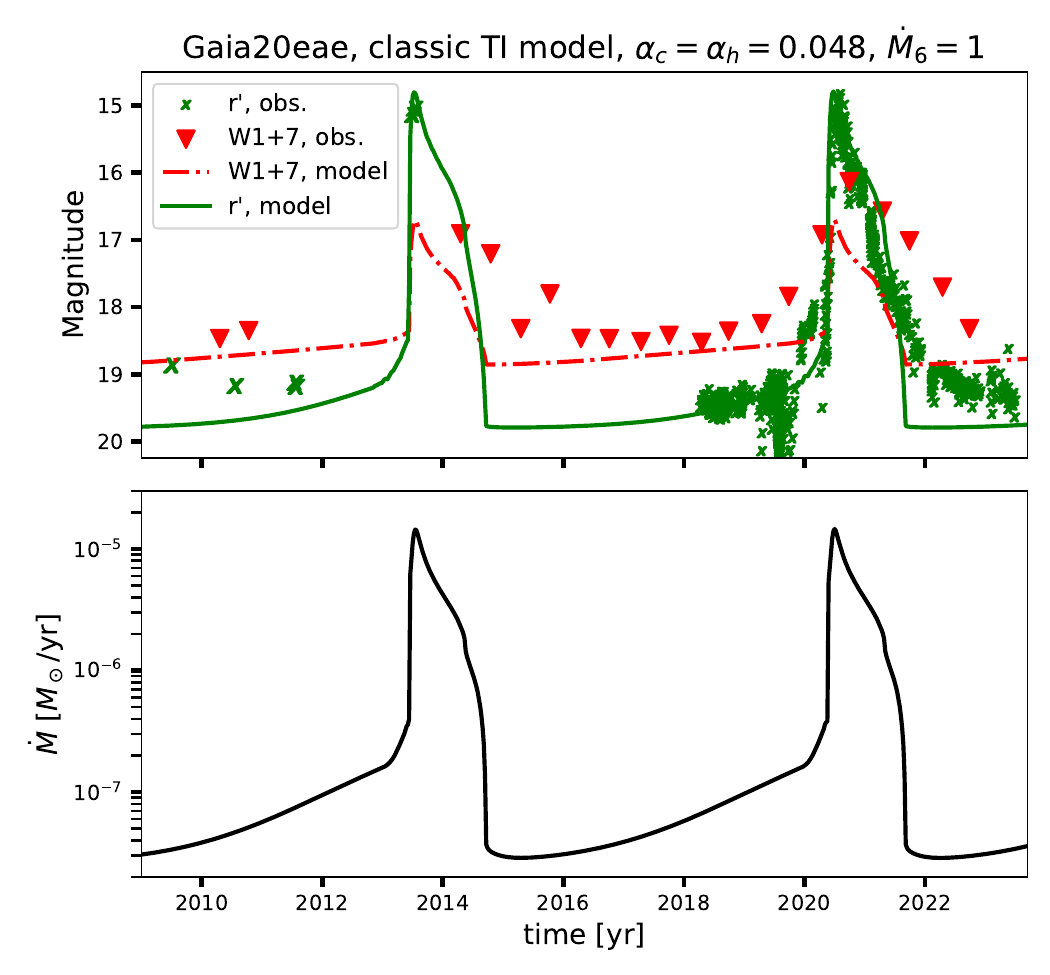}
\caption{Top: Sloan $r'$ and WISE W1 bands data for the outburst of Gaia20eae \citep{CruzSaenz-22-Gaia20eae} are shown with triangles. The lines show our model disc magnitudes in these two bands. Bottom: Stellar accretion rate history for this disc model.}
\label{fig:Gaia20eae_classic_TI}
\end{figure}

Gaia20eae \citep{CruzSaenz-22-Gaia20eae} is another curious case of a YSO outburst that does not easily fall into the FUOR/EXOR dichotomy. This source drew attention through the {\em Gaia} alert system when its brightness increased by just over 4 mag in 2020. \cite{CruzSaenz-22-Gaia20eae} observed the source in the optical and NIR and classified it as an EXOR-like young eruptive star. Fig. \ref{fig:Gaia20eae_classic_TI} shows the Sloan filter $r'$ and IR W1 band photometry data\footnote{For brevity purposes in Fig. \ref{fig:Classic_TI_Sigma_T} we combined the r' band photometry data from ZTF and the RC80 telescope.} for the source with open triangles, along with our model that we discuss below. Gaia20eae distance and extinction are estimated at $D = 2.8$~kpc and $A_{\rm V} = 4.7\pm 0.8$ mag. Pre-outburst spectra of the source show that this is a Class II YSO with $T_{\rm eff} = 4,300\pm 300$~K, radius $R_* = 3.3\, \rsun$ and mass $M_* = 1.15\,\msun$. 

Gaia20eae displayed a number of EXOR-like traits \citep{CruzSaenz-22-Gaia20eae}: (i) The outburst rise time and duration are comparable but a factor of $\sim 2$ longer than that of major outbursts of EX Lup \citep[e.g.,][]{Wang-23-EXLUP,Kospal23-EXLup}; (ii) its spectra show strong emission lines; (iii) the outbursts could be recurrent, with a repetition time of about $\sim 7$ years. 

On the other hand, (i) the peak bolometric luminosity of Gaia20eae is over $100\, \lsun$, well within what is typical for FUORS \citep{Fischer-PPVII}; (ii) the accretion rate in quiescence and at the peak of the outburst are $\sim 10^{-7}$ \MSunPerYear and $1.6\times 10^{-5}$ \MSunPerYear, respectively, both of which are quite typical of FUORs. For comparison, the accretion rate in EX Lup during the brightest of the bursts is between $\dot M \sim 10^{-7}$ \MSunPerYear \citep{Wang-23-EXLUP} 
and $\sim 10^{-6}$ \MSunPerYear \citep{Cruz-23-EXLup}\footnote{The discrepancy is due to the uncertainty in extinction towards EX Lup, see Table \ref{tab:Table2}.};
(iii) Further, high resolution spectra capable of detecting absoption lines were carried out in Gaia20eae $\sim 7$ months after the source started its rapid decline from the maximum light \citep[see Fig. 3 in][]{CruzSaenz-22-Gaia20eae}. By that time the source was almost 10 times dimmer than at its peak. It is thus potentially possible that while Gaia20eae spectrum was EXOR-like at the time of the observation, it was more FUOR-like at the maximum light.

We tried to find a classic TI model capable of accounting for the main characteristics of Gaia20eae. Similar to the process described in \S \ref{sec:PTF14jg}, we tested a couple dozen model parameter combinations for $\alpha_{\rm c}$, $\alpha_{\rm h}$, $\dot M_{\rm feed}$, while keeping $T_{\rm c} = 8,000$~K to avoid reflares that do not appear to be observed in this source. We set stellar parameters to the same values as those derived by \cite{CruzSaenz-22-Gaia20eae}. The model with $\alpha_{\rm c} = \alpha_{\rm h} = 0.048$, with $\dot M_{\rm feed} = 10^{-6}$\MSunPerYear worked best. This model is shown in Fig. \ref{fig:Gaia20eae_classic_TI} with lines. The top panel shows the r' and W1 band photometry 
of Gaia20eae with symbols, and that of our model disc with lines, assuming the distance of 2.8 kpc, $A_{\rm V} = 4.3$ mag, and disc inclination of 60$^\circ$. The bottom panel in Fig. \ref{fig:Gaia20eae_classic_TI} shows the respective accretion rate onto the star.

We see that the model $\dot M$ during the outburst is close to the values derived by \cite{CruzSaenz-22-Gaia20eae}, and the simulated lightcurve is qualitatively similar to that observed in Gaia20eae. On the other hand, the model is not sufficiently bright in the W1 band. Also, the model outburst starts too late and ends too early. It is possible that a higher feeding rate would remedy some of these model shortcomings as the disc would be brighter; the size of the TI-active region, $R_{\rm TI}$, would also be larger, producing longer outbursts. We plan to make a wider parameter space exploration in a follow up paper. 

\bref{One technical note is in order. It interesting that outbursts with $\alpha_{\rm c} = \alpha_{\rm h}$ are quite weak in dwarf novae systems, which is one of the main reasons in requiring $\alpha_{\rm h}$ to usually exceed $\alpha_{\rm c}$ strongly \citep[e.g.,][]{Lasota01-Review,Hameury-20-review}. In Fig. \ref{fig:Gaia20eae_classic_TI} we however see quite a strong TI outburst.  We recomputed our S-curves for a dwarf novae case, and investigated the sensitivity of the outburst magnitude, defined as the ratio of $\dot M$ at the peak of the burst to that just before an outburst, to the $\alpha_{\rm h}/\alpha_{\rm c}$ ratio. We confirm  what is well known in the dwarf novae field: outbursts become very weak when the ratio approaches unity (we used $\dot M_{\rm feed} =  10^{-10}$\MSunPerYear for these experiments). For FUOR parameters, however, typical disc radii are much larger, and disc densities are much lower, and this leads to significant opacity changes around the opacity gap \citep[e.g., Fig. 9a in][]{Bell94}. These differences lead to corresponding differences in the S-curves, and hence outburst magnitude seem to be less sensitive to $\alpha_{\rm h}/\alpha_{\rm c}$ in the protoplanetary disc case, although there is still some dependence that can be seen when the two top panels in Fig. \ref{fig:Lots_of_bursts} are compared. Also, viscosity parameters assumed in these experiments remain instrumental in controlling outburst duration and detail of the lightcurve shape.}

\subsection{Gaia21bty and Gaia19bey}\label{sec:Gaia21bty}

Another intriguing source straddling the boundary between EXOR and FUOR classifications is Gaia21bty \citep{Siwak-23-Gaia-21bty}. The authors find that the source exhibits a lightcurve of an EXOR, e.g., the brightest part of the outburst lasts $\sim (4-6)$ months. At the same time, Gaia21bty displays FUOR-like spectra, and the estimated stellar accretion rate at the peak of the burst, $\dot M_{\rm peak} \sim 2.5 \times 10^{-5}$\MSunPerYear, is typical of FUORs. Gaia21bty shows multiple re-brightening events in {\em Gaia} G band after the decline from the highest peak \citep[see Fig. 1 in][]{Siwak-23-Gaia-21bty}. The events appear to last a few months or so, not unlike the secondary peak for PTF14jg.

There is a sparser time and wavelength coverage of this enigmatic source, and therefore we do not attempt a detailed source modelling at this time. However, we point out that the observed properties of Gaia21bty are generally consistent with the classic TI bursts, and the secondary smaller outbursts could be due to reflares.


Finally, a burster worth mentioning as peculiar is Gaia19bey \citep{Hodapp20-Gaia19bey}. The {\em Gaia} G band lightcurve of the source shows a relatively smooth rise and then a plateau $\sim 2$ magnitudes above the pre-burst photometric point. This phase lasts $\sim 3$ years, but then there is a $\sim 2$ magnitudes further increase in brightness and a fall from it back to pre-outburst brightness within less than a year \citep[Fig. 2 in][]{Hodapp20-Gaia19bey}. This timescale is consistent with a classic TI outburst. The smooth rise of the flux during 2015-2016 to a plateau later on is somewhat reminiscent of the TI model with \cite{ZhuEtal09} opacity shown in Fig. \ref{fig:Comp_Zhu_to_BL94}. Gaia19bey also showed re-brightening events during the declining phase of the burst, with the largest of them having about 1 magnitude and a timescale of about 50 days. These secondary re-brightening events are somewhat shorter and less powerful than those shown in the models in Figs. \ref{fig:PTF14_D2} and \ref{fig:PTF14_D5} but could nevertheless be related to reflares during classic TI bursts. Finally, \cite{Hodapp20-Gaia19bey} point out that there is evidence for a powerful prior outburst from this source from the 2MASS survey some $\sim 20$ years prior. This repetition time scale is within the range expected from the classical TI outbursts (eq. \ref{t_rep}).

\section{Discussion}\label{sec:discussion}

In this paper we summarised the properties of disc accretion outbursts due to the classic Thermal Instability of protoplanetary discs assuming modern disc viscosity formulation that is broadly consistent with observations of TI in dwarf novae and low mass X-ray binaries, and with modern 3D numerical simulations (see Introduction). These viscosity values are about two orders of magnitude larger than those used by the previous workers \citep{Clarke-90-FUOR,Bell94,BellEtal95,LodatoClarke04,Clarke05-FUORs} who aimed to account for the then known classic FUORs, only a few in number at the time. We discussed the ingredients  and uncertainties of our time-dependent 1D disc calculations of TI (\S \ref{sec:methods}),  showed examples of classic TI outburst lightcurves, disc evolution during the instability cycle, and approximate scalings of outburst properties with model parameters (\S \ref{sec:general}). 

In \S \ref{sec:Unusual_fuors} we showed that classic TI provides a promising framework to understand several  recently observed episodic accretion bursters with unusual properties, PTF14jg \citep{Hillenbrand19-PTF14}, Gaia20eae \citep{CruzSaenz-22-Gaia20eae}, Gaia21bty \citep{Siwak-23-Gaia-21bty} and Gaia19bey \citep{Hodapp20-Gaia19bey}: the luminosity and some other properties typical of FUORs but surprisingly short duration of a year to a few, more consistent with the EXOR class \citep{AudardEtal14,Fischer-PPVII}.

We now discuss how these results fit within our current understanding of episodic accretion in young accreting stars.

\subsection{The role of Thermal Instability in FUORs}\label{sec:Discussion_TI_in_FUORs}

\subsubsection{Burst onset}\label{sec:burst_onset}

We propose that both classic (such as FU Ori) and the unusually short FUOR outbursts {\em begin} with an upward transition from the cold stable branch of the S-curve towards the hot stable branch in the inner $\sim 0.1$~au. This proposal has a number of attractive features:

\begin{enumerate}
    \item FUORs have quiescent stellar accretion rates of $\dot M\sim (10^{-9} - 10^{-7})$\MSunPerYear \citep{HK96,AudardEtal14}. This corresponds rather well to $\dot M$ on the cold stable branch of the S-curve at a few stellar radii, cf. the left panel of Fig. \ref{fig:S-Curve}. This region is relevant here because this is where the classic TI is triggered (Fig. \ref{fig:Classic_TI_Sigma_T}). Furthermore, even in scenarios where TI is triggered at larger radii, e.g, $R\sim 10 \rsun$ as in Fig. 4 in \cite{LodatoClarke04} or $R\sim 20\rsun$ in Fig. 10 in \cite{Nayakshin23-FUORi-2}, the pre-outburst $\dot M$ must still be  limited by the maximum stable $\dot M$ on the low branch at a few stellar radii -- or else the burst would have been triggered there rather than further out.     
    
    \item The TI outburst rise time is of the order of the ionisation front propagation time through the unstable region, which is a fraction of a year. In the case of the classic TI, the trigger occurs at small radii and the front propagates outward; eq. \ref{t_rise} yields months for the rise time. As found by \cite[][]{LodatoClarke04,Nayakshin23-FUORi-2}, the TI-unstable region may be somewhat larger if there is a massive planet embedded in the inner $\sim 0.1$ au of the disc. The burst can then start behind the planet and propagate inward, which may increase the burst rise time a little:
    \begin{equation}
    t_{\rm rise} \sim  \frac{R}{\alpha_{\rm h} c_{\rm h}} = \frac{P_{\rm orb}(R)}{2\pi \alpha_{\rm h} h}
    \sim 0.5 {\rm \, yr \,} \alpha_{-1}^{-1} h_{-1}^{-1} \, R_{-1}^{3/2}\;
    \end{equation}
    where $P_{\rm orb}(R)$ is the orbital period at the trigger radius, $R$, $R_{-1} = R/(0.1$au), $h = H/R$ on the hot branch, $h_{-1} = h/(0.1)\sim 1$, and $\alpha_{-1} = \alpha_{\rm h}/(0.1) \sim 1$. We see that in both scenarios the burst rise time is comparable with that of many FUORs. Multiwaveband observations can pinpoint the location of the outburst trigger, see \S 6.2 in \cite{Liu22-FUORs} and \S 5 in \cite{Nayakshin23-FUORi-2}. Alternative models that do not appeal to TI and are triggered at, or bring material from, much larger distances produce rise times much longer than a year. For example, in the scenario of \cite{ArmitageEtal01} the outburst is triggered at $\gtrsim 1$~au, where just the orbital time is already longer than a year, and the rise time is thus guaranteed to exceed that many fold. While this issue has not been directly addressed in the literature, 2D simulations of \cite{Kadam20-DZ-MRI,Vorobyov-21-flyby-FUORs} show MRI activation burst rise times of at least tens of years. On the other hand, MRI bursts started by activation of TI in the inner disc may have rise times commensurate with observations.

    \item Model TI outbursts yield peak (maximum) burst accretion rates in the rough range  $(10^{-5} - 10^{-4})$\MSunPerYear (eq. \ref{dotM_peak}), which is generally consistent with $\dot M_{\rm peak}$ inferred in well characterised FUOR outbursts \citep[e.g.,][]{ZhuEtal07,Kospal-16-HBC722,Szabo21-V1057,Szabo22-V1515,2022Lykou,Carvalho-23-V960}. See \S \ref{sec:discussion_burst_desrt} for a further discussion of this.
\end{enumerate}

\subsubsection{What powers the long (classic FUOR) bursts?}\label{sec:discussion_burst_powering}

On the other hand, as is well known \citep{Bell94}, the TI-unstable disc region is very small ($R\lesssim 0.1$~au, eq. \ref{R_TI}). With the modern disc viscosity values, $\alpha_{\rm c}\gtrsim 0.01$, this region contains some tens of Earth masses of gas only (eq. \ref{M_TI}), which is enough to sustain outbursts at the typical FUOR $\dot M$ for just a few years. The classic TI may thus account for short 
unusual FUORs such as PTF14jg and Gaia20eae but not for the more prototypical long bursts that last tens to hundreds of years \citep[e.g., see Table A1 compiled by][]{Vorobyov-21-flyby-FUORs}.

One plausible solution is that while TI triggers a FUOR outburst, some other source of matter continues to power longer outbursts such as the one in FU Ori. This is the situation in the scenario proposed by \cite{Nayakshin-23-FUOR,Nayakshin23-FUORi-2}, where a massive ($\sim 5 \mj$ or more) planet migrates into the inner disc. While the planet is far from the TI unstable region, it is embedded into relatively cool regions of the discs, $T_{\rm c} \lesssim 2,000$~K. These cool regions are always on the cold stable branch of the S-curve. However, when the planet migrates inward of $\sim 0.1$~au, it is exposed to disc temperatures as high as $30,000$~K during TI outbursts. If the planet is sufficiently fluffy then its envelope begins to evaporate and this provides an extra fuel supply that keeps the inner disc at the hot stable S-curve branch for much longer than expected in planet-free discs.


Further theoretical work should aim to establish whether a source of additional mass other than a massive planet can satisfy constraints from the disc physics and the surprisingly tight constraints on the active disc radius of 0.3 au in FU Ori \citep{2022Lykou}. From the observational perspective, interferometry of FUOR discs, both in and post-outburst, with instruments such as MATISSE and BIFROST proposed by \cite{Kraus-22-BIFROST} may constrain the presence of massive planets in FUOR discs. Photometric and emission/absorption line variability observations of FUORs is another valuable tool \citep[e.g.,][]{PowellEtal12,Siwak21-FUOri-QPOs} to constrain planet presence.

There may also be statistical tests of planet presence in the innermost regions of protoplanetary discs undergoing FUOR/EXOR type variability. Since planets modify the discs by opening deep gaps and even inner holes \citep[e.g.,][]{LodatoClarke04}, accretion variability of planet-bearing and planet-free discs may have different statistics and trends with system parameters, such as stellar mass, and system age/class. While observational statistics on the frequency of eruptive YSOs is patchy \citep[e.g.,][]{AudardEtal14}, it is gradually improving \citep[e.g.,][]{Contreras-19-FUOR-statistics,Fischer-19-FUORs-Orion,Contreras-Pena-23-SixFUORs}.

\subsection{The TI burst desert: a statistical test of TI}\label{sec:discussion_burst_desrt}


Local vertical thermal balance calculations show that for any disc annulus there exists a range in feeding accretion rates, $\dot M_{\rm max} < \dot M_{\rm feed} < \dot M_{\rm min}$,  where this annulus is thermally unstable (cf. Fig. \ref{fig:S-Curve} and \S \ref{sec:scaling}). The unstable range is a strong function of radius ($\dot M\propto R^{2.4}$, eqs. \ref{S-curve_scaling}), however non-local disc models show that the inner TI-active portion of the disc acts very much in unison to display globally coherent instability cycles \citep[e.g.,][]{Faulkner83-DNe}. Further, the instability must be quenched when the unstable region ($R_{\rm TI}$, eq. \ref{R_TI}) is just slightly larger than the star. This sets a well defined $\dot M_{\rm crit}$ below which all radii in the disc are stably accreting on the cold branch of the S-curve. \cite{Bell94} find that TI is triggered only in discs with $\dot M_{\rm feed} > \dot M_{\rm crit} \sim$ a few$\times 10^{-7}$\MSunPerYear. Here we used viscosity parameter values two orders of magnitude higher, however obtained similar results, which is probably due to $\dot M_{\rm crit}$ being independent of $\alpha_{\rm c}$ to the zeroth order (eq. \ref{dotM_crit}). In \S \ref{sec:param_space} we considered approximate scaling of $\dot M_{\rm crit}$ with parameters of the system, and concluded that it may be expected to be a relatively weak function of $M_*$. 

This weak dependence of $\dot M_{\rm crit}$ on $\alpha$ parameter and $M_*$ makes for an interesting observationally testable prediction of TI presence in protoplanetary discs. Discs with $\dot M_{\rm feed} > \dot M_{\rm crit}$ must switch from the low stable to the hot stable branch of the S-curve, with $\dot M$ suddenly increasing by $\sim(2-3)$ orders of magnitude. While during the outburst $\dot M$ decreases  from its peak value (e.g., Fig. \ref{fig:Lots_of_bursts}), simulations suggest that at the end of the outburst $\dot M$ is still an order of magnitude or so larger than it is before the outburst. We therefore expect a dearth (desert) of discs with $\dot M$ just above $\dot M_{\rm crit}$. The desert is expect to be (1-2) orders of magnitude wide in $\dot M$.


It may appear other disc instabilities or actors may dilute and/or wash away the TI burst desert by influencing $\dot M_{\rm feed}$. To name a few most relevant to consider: the dead zone ionisation instability \citep[][]{ArmitageEtal01,Zhu10-DZ-MRI-1D}; \cite{Toomre64} disc self-gravity instability due to which massive discs can drive strong spiral density waves \citep{RicciEtal15,KratterL16} and even fragment onto planet-mass objects \citep{VB06,VB15,Vorobyov-Elbakyan-19}. The inner discs can also be strongly influenced by binary interactions \citep{Bonnell92-binary-FUORs}, stellar fly-bys or clump infall \citep{Vorobyov-21-flyby-FUORs,Borchert-22-flyby-FUORs}.

However, all of the phenomena listed above affect the disc at large, $R\gtrsim 1$~au, radii first, and only then disc physics propagates them to much smaller scales. The classic TI is triggered very close to the star, $R\sim 2R_*$ \citep[Fig. \ref{fig:Classic_TI_Sigma_T}, see also][]{Bell94}. In fact, its full domain of operation is only $R_* < R < R_{\rm TI}\lesssim 0.1$~au (eq. \ref{R_TI}). Therefore, the larger scale effects may be instrumental in dictating the mass supply rate ($\dot M_{\rm feed}$) into the innermost disc, but the physics of TI still forces it to be either on the cold or the hot stable branch of the S-curve at any given moment. For example, in models that include both the dead zone ionisation instability and the TI, multiple powerful TI outbursts occur on the top of a much longer dead zone ionisation instability outburst  when one models the accretion flow all the way to the star \citep[cf. \S 5.3 and the Appendixes in][]{Nayakshin23-FUORi-2}. In this example,  the dead zone ionisation instability controls $\dot M_{\rm feed}$ on timescales $\gtrsim 10^3$ while TI controls $\dot M$ onto the star, with TI cycle repetition time of a decade or so. In other words, instabilities of protoplanetary discs at $R \gg 0.1$~au scales cannot turn off TI in the inner disc.

This prediction of a desert in $\dot M$ in accreting YSO applies to both quiescent and burst/outburst phases of episodic accretors whether they are classified as FUORs, EXORs, or a combination of thereof. For a preliminary observational test of  the burst desert we focus on the maximum burst accretion rate, $\dot M_{\rm burst}$, measured at peak light. 
We compiled a list of publications for accreting YSOs that derived $\dot M_{\rm burst}$ for EXOR- and FUOR-type sources, including those with mixed characteristics. The result is listed in Table \ref{tab:Table2}. For FUORs, the accretion rate onto the star during outbursts is usually constrained via fitting the observed SED with the steady state accretion disc model \citep[e.g.,][]{HK96,Liu22-FUORs,Rodriguez-22-FUORs}. For EXORs, deriving $\dot M$ by this method is not appropriate as their spectra are usually dominated by the radiation from the star and its accretion hot spot \citep[e.g.,][]{Cruz-23-EXLup,Wang-23-EXLUP}, so one needs to employ a model for the accretion shock. In both cases, an accurate value for the distance to the source is crucial. Therefore, when several publications with estimates of $\dot M$ exist for the same source we select those with the most recent distance determination, which is usually the Gaia-based parallax distance measurement. For some of the sources, e.g., V1118 Ori and EX Lupi, there is a significant uncertainty in the extinction, which drives a large uncertainty in $\dot M_{\rm burst}$. Further, we broke the list of the sources in two groups: those that we believe are on the cold or the hot branch of the S-curve at the maximum light of their bursts, respectively. This grouping of course has no effect on the resulting distribution of $\dot M_{\rm burst}$. The sources  are listed in Table \ref{tab:Table2} in the order of increasing $\dot M_{\rm burst}$ for convenience. 

\begin{figure}
\includegraphics[width=1\columnwidth]{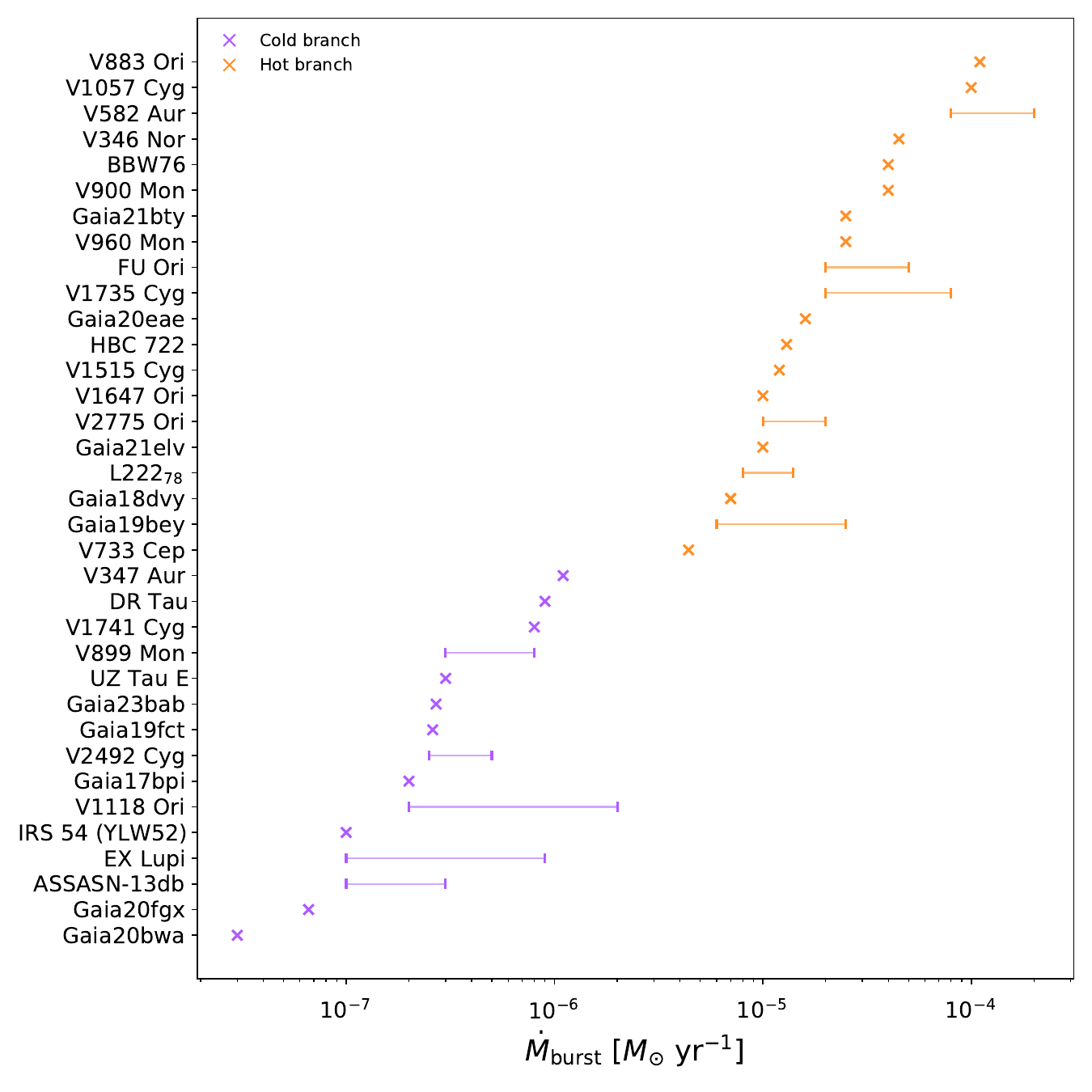}
\caption{The maximum accretion rate onto the star during eruptions of episodic accretors with large increases in the optical, cf. Table \ref{tab:Table2} for indivudual sources and references. The compilation of sources includes both EXORs and FUORs, and the sources with mixed characteristics. The data show that there is a deficit (desert) of bursts with $\dot M_{\rm burst}$ in the range between $\sim 10^{-6}$ and $10^{-5}$\MSunPerYear. Classic TI predicts such a desert (see \S \ref{sec:discussion_burst_desrt}).}
\label{fig:mdot_statistics}
\end{figure}

\begin{figure}
\includegraphics[width=1\columnwidth]{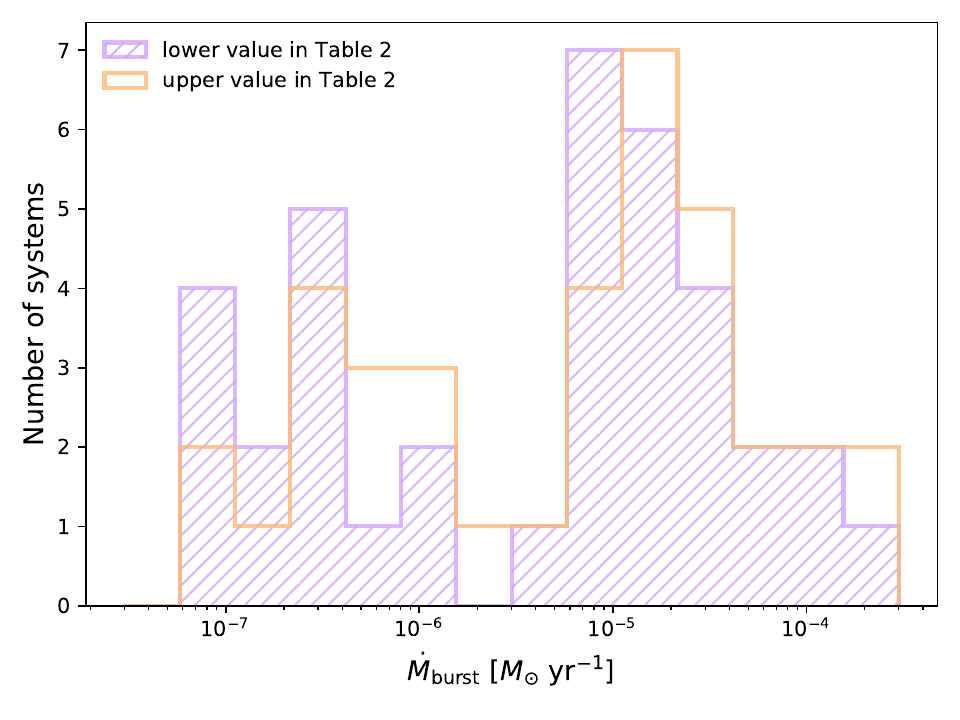}
\caption{Same data as in figure \ref{fig:mdot_statistics} but shown as a histogram. The shaded (open) histogram shows the data in Table \ref{tab:Table2} with $\dot M_{\rm burst}$ taken at the lowest (highest) value for the sources where more than one value for $\dot M_{\rm burst}$ is found in the literature. The `desert' is clearly seen here between $\sim 10^{-6}$ and $10^{-5}$\MSunPerYear.}
\label{fig:mdot_histogram}
\end{figure}

Figures \ref{fig:mdot_statistics} and \ref{fig:mdot_histogram} show the data from Table \ref{tab:Table2}. We see that there is indeed a suggestive gap in $\dot M_{\rm burst}$ between sources with $\dot M_{\rm burst} \lesssim 10^{-6}$\MSunPerYear and $\dot M_{\rm burst} \gtrsim 10^{-5}$\MSunPerYear. The errors involved in deriving $\dot M_{\rm max}$ probably have a broad Gaussian distribution in $\log \dot M_{\rm burst}$, which must serve to dilute any sharp features present in the actual distribution. For this reason\footnote{Additionally, the exact location of the gap in the accretion rate depends on many system parameters, i.e., disc opacity, radius of the star (\S \ref{sec:scaling}) and, potentially, stellar magnetic fields, which we neglected here. This is observationally justifiable for FUORs as their discs are believed to persist all the way to the star \citep[e.g.,][]{HK96,AudardEtal14}, but at $\dot M$ appropriate to EXORs we expect an inner magnetospheric disc cavity \citep[e.g.,][]{Hartmann16-review}. Stars with strong stellar magnetic fields may be expected to have a higher $\dot M_{\rm crit}$ threshold for TI due to the existence of this cavity. Accounting for these modelling uncertainties should serve to wash out any sharp features in the distribution of $\dot M_{\rm burst}$. The fact that the desert is very clear in the data is thus very significant.} it is likely that the actual (error-free) histogram of $\dot M_{\rm burst}$ has a yet drier desert than that seen in Fig. \ref{fig:mdot_histogram}.

Furthermore, we are not aware of an observational bias that would account for the gap, or  disc physics other than TI that would predict the desert where it is observed. For example, the dead zone ionisation instability \citep{ArmitageEtal01} predicts that $\dot M\sim 10^{-8}$\MSunPerYear in quiescence, and so should lead to a desert at much lower values of $\dot M_{\rm burst}$. \bref{Additionally, \cite{Bae14-MRI-2D} show in their Fig. 16 that this instability predicts that sources should spend a significant fraction of time in a quasi-stable accretion mode with $\dot M$ slightly higher than $10^{-6}$\MSunPerYear. This contradicts the data shown in Fig. \ref{fig:mdot_histogram}. However, 2D simulations by \cite{Bae14-MRI-2D} were computationally more expensive than our 1D disc modelling, so they excluded the disc region within 0.2 AU of the star. This is exactly where TI operates. It is likely that had the authors been able to model this region, then TI would operate and would have erased the quasi-steady accretion island in their Fig. 16. This would possibly transforme their $\dot M$ distribution into something much more consistent with what we find here.} We therefore preliminary conclude that the data hint strongly that TI is active in {\em all} episodically accreting YSOs. We plan to firm this conclusion up with population synthesis of episodically accreting YSOs in the future.



\begin{table*}
    \centering
    \begin{tabular}{r|ccl}
         Source &    $\dot M_{\rm burst}$ [$M_\odot$ yr$^{-1}$]  & $M_*$ [$\msun$] & Reference\\
         \hline
         \hline
         Cold branch: &  & & \\
         \hline
         Gaia20bwa & $3\times 10^{-8}$ & 0.28 &  \cite{Nagy22-two-EXORs}\\
         Gaia20fgx & $6.6\times 10^{-8}$ & 0.53 &  \cite{Nagy22-two-EXORs}\\
        IRS 54 (YLW52) & $\sim 10^{-7}$ & (0.1-0.2)  &\cite{Stock20-EXori}\\
         ASSASN-13db & $(1-3)\times 10^{-7}$ & 0.15 &  \cite{Sicilia-Aguilar-17-ASSASN-13db}\\
         Gaia17bpi & $2 \times 10^{-7}$ & 0.63 & \cite{Rodriguez-22-FUORs}\\
         V2492 Cyg & $(2.5-5)\times 10^{-7}$ & 0.7 & \cite{Kospal13-V2492,Giannini18-V2492}\\
         Gaia19fct & $2.6\times 10^{-7}$ & 0.44 & \cite{Park22-Gaia19fct}\\
          Gaia23bab & $ 2.7\times 10^{-7}$ & ? & Giannini et al. (submitted)\\
         V1118 Ori & $(2-20)\times 10^{-7}$ & 0.29 &\cite{Giannini17-V1118-exor}\\
         UZ Tau E & $3\times 10^{-7}$ & ? &  \cite{LorenzettiEtal09}\\
         V899 Mon & $(3-8)\times 10^{-7}$ & (0.5-5) & \cite{Ninan-15-V899} \\
        V1741 Sgr & $8\times 10^{-7}$ & 0.7 & \cite{Kuhn-23-V1741}\\
         DR Tau  & $9\times 10^{-7}$ & ? &  \cite{LorenzettiEtal09}\\
         EX Lupi & $\sim (1-9)\times 10^{-7}$ & 0.83  &\cite{Wang-23-EXLUP,Cruz-23-EXLup}\\
         V347 Aur & $1.1 \times 10^{-6}$ & 0.35 & \cite{Dahm20-V347}\\
    \hline
    \hline
         Hot branch: &  &  & \\
         \hline
         V733 Cep & $ 4.4\times 10^{-6}$ & 0.5 &  Park et al., in prep. \\
         Gaia18dvy & $\sim 7\times 10^{-6}$ & 1.0 & \cite{Szegedi-Elek-20-Gaia-18}\\
         Gaia19bey & $(0.6-2.5)\times 10^{-5}$ & ? & \cite{Hodapp20-Gaia19bey}\\
         Gaia21elv & $\sim 1\times 10^{-5}$ & 1.0 & \cite{Nagy23-Gaia21elv}\\
         L222$_{-}$78 & $(0.8-1.4) \times 10^{-5}$ & 1.0 & \cite{Guo24-L222} \\
         V1647 Ori & $\sim 10^{-5}$ & 0.5 &\cite{Muzerolle05-V1647}\\
         V1515 Cyg & $1.2\times 10^{-5}$ & 0.3 & \cite{Szabo22-V1515}\\
         V2775 Ori & $(1-2)\times 10^{-5}$ & 0.24 &\cite{Fischer12-V2775}\\
         HBC 722 & $1.3\times 10^{-5}$ & 0.66 &\cite{Rodriguez-22-FUORs}\\
         Gaia20eae & $1.6\times 10^{-5}$ & 1.15 & \cite{CruzSaenz-22-Gaia20eae}\\
         V1735 Cyg & $(2-8)\times 10^{-5}$ & 1.0 & Szab\'o et al., in prep.\\
         FU Ori & $ (2-5)\times 10^{-5}$ & 0.6 &\cite{2022Lykou}\\
         Gaia21bty & $ 2.5\times 10^{-5}$ & 0.23 & \cite{Siwak-23-Gaia-21bty}\\
         V960 Mon & $ 2.5\times 10^{-5}$ & 0.59 & \cite{Carvalho-23-V960}\\
         BBW76 & $ 4\times 10^{-5}$ & 0.8 & \cite{Siwak-20-BBW76}\\
         V900 Mon & $ 4\times 10^{-5}$ & 1.0 & \cite{Lykou24-V900Mon}\\
         V346 Nor & $ 4.5\times 10^{-5}$ & 1.0 &\cite{Kospal-17-V346,Kospal-20-V346}\\
         V1057 Cyg & $\sim 10^{-4}$ & 1.0 & \cite{Szabo21-V1057}\\
         V883 Ori & $1.1\times 10^{-4}$ & 1.3 & \cite{Liu22-FUORs}\\
         V582 Aur & $ (0.8-2)\times 10^{-4}$ & 1.0 & \cite{Zsidi-19-V582}\\
         RNO 54 & $ 3 \times 10^{-4}$ & 0.23 & \cite{Hillenbrand23-RNO54}\\
         \hline
    \end{tabular}
    \caption{A compilation of published values for the maximum accretion rate onto the star, $\dot M_{\rm burst}$, during eruptions of episodic accretors with large increases in the optical. Deduced or assumed values for the stellar mass are also given when available. See \S \ref{sec:discussion_burst_desrt} for detail.}
    \label{tab:Table2}
\end{table*}

\subsection{FUOR occurrence rates}\label{sec:FUOR_statistics}

\bref{While the first draft of this paper was in review, new results from the over decade long wide sky VVV and VVVX NIR surveys of class I YSOs \citep{Guo24-VVV} were published, with over a dozen new eruptive variables identified and confirmed as FUORs. \cite{Contreras-Pena-24} classified FUOR-type outbursts into short (duration less than 1 year), intermediate, and long (duration greater than 9 years). In the context of our model, the short outbursts are most like those produced by the classic TI scenario. \cite{Contreras-Pena-24} conclude that the fraction of their sources showing short outbursts is $\sim 1$\%. It is interesting to compare these results with theoretical expectations. While a detailed comparison is beyond the scope of the present paper and will require a population synthesis study coupled with consideration of selection effects, some preliminary qualitative conclusions can be made based on Fig. \ref{fig:Lots_of_bursts}. Defining outburst duration as time when $\dot M\gtrsim $ a few $\times 10^{-6}$\MSunPerYear, we conclude that outburst duty cycle is roughly 10\%. What is more, it appears that an observing time window with a duration of $\sim 10$ years, thrown randomly across the horisontal axis in Fig. \ref{fig:Lots_of_bursts} is quite likely to catch one of the bursts. In other words, qualitatively, our models predict that a very high, $\sim 50$\%, fraction of YSO discs should experience a burst during an interval of 10 years. This is clearly much larger than the fraction found by the VVV/VVVX survey in \cite{Contreras-Pena-24}. 

This disagreement between the classic TI scenario and short FUOR outbursts statistics may well be yet another example indicating that TI is only a part of the story in episodic accretion. It appears unlikely that $\dot M_{\rm feed}$ is lower than the instability threshold (a few $\times 10^{-7}$\MSunPerYear) in $\sim 98$\% of the VVV/VVVX sample. We would rather argue that an agent -- potentially the same one as needed for the long FUORs -- is suppressing the supply of matter in the innermost 0.1 au in most of the sample. For example, massive planets are known to open wide deep gaps in protoplanetary discs, now directly observable with ALMA \citep[e.g.,][]{Mesa19-PDS70,Isella19-PDS70-ALMA}. It is possible that many of the discs in the \cite{Guo24-VVV} sample have inner holes opened by massive planets, albeit on much smaller scales, $R\lesssim 1$~au. Another scenario could be dead zones \citep{ArmitageEtal01} stifling accretion into the inner disc for long periods of time, whilst they accumulate enough mass to be released during the long FUOR outbursts.}

\subsection{The global role of TI in the episodic accretion phenomenon}\label{sec:discussion_global}

In \S \ref{sec:Discussion_TI_in_FUORs} we argued that TI provides a promising framework for understanding of the intermediate type FUOR/EXOR bursters, and the beginning phases of classic (long) FUORs. At the same time, long FUORs require too high mass budget for the classic TI to power them (\S \ref{sec:discussion_burst_powering}), so some other agent needs to provide that mass.

An attractive hypothesis is that TI is the physics that divides the episodic accretors into the two major classes. Historically,  EXOR and FUOR classes were believed to be rather well separated in terms of their spectral and time variability properties \citep[e.g.,][]{Herbig-77-FUOR-EXOR,HK96,AudardEtal14}. Observations of the last decade or so show that there is more of a continuum change between the two classes \citep[e.g.,][]{Fischer-PPVII}. Nevertheless, if we focus on the maximum accretion rates during the burst, then there is still a significant separation between the classes. Most of the sources in the lower part of Table \ref{tab:Table2} are classified as EXORs, and their $\dot M_{\rm burst} \lesssim 10^{-6}$\MSunPerYear, by a significant factor for some of the bursters. It is possible that EXORs are the sources that remain on the cold branch of the S-curve, even during bursts. Their eruptions should then be driven by something else, e.g., instabilities connected with the interaction of stellar magnetic fields with the inner disc \citep[e.g.,][]{DAngelo-12-EXORs,Armitage16-EXORs}. At the same time, most of the sources in the second part of Table \ref{tab:Table2} are FUORs.


That there exists a desert in $\dot M_{\rm burst}$ between the two classes of episodic low mass YSO accretors demands an explanation, and the classic TI is an excellent candidate for it.
TI outbursts provide a successful framework for understanding of accretion outbursts of related systems -- DNe and LMXRBs (see Introduction). It should not be surprising that episodic accretion in YSO also requires TI as an important ingredient. However, additional physics/effects not operating in DNe and LMXRBs must be present in the discs of rapidly accreting YSOs to account for the full spectrum of their episodic large magnitude variability and statistics (\S \ref{sec:FUOR_statistics}.

\section{Conclusions}

In this paper we showed that classic TI with modern viscosity choices (a) provides an attractive framework for large magnitude bursts with characteristics intermediate between FUORs and EXORs; (b) accounts well for outburst rise time and maximum brightness in long (classic) FUORs; (c) predicts that there should exist a dearth/desert of outbursts with $\dot M_{\rm burst}$ between $\sim 10^{-6}$\MSunPerYear and $\sim 10^{-5}$\MSunPerYear, and that published data provide preliminary support to this. Classic EXORs (FUORs) may be sources that are on the cold (hot) stable branch of the S-curve during their bursts (outbursts), so TI may be instrumental to the exstense of the two long established classes of episodically erupting YSOs. At the same time, TI fails to account for a number of important characteristics of the episodic accretion phenomenon (\S \ref{sec:discussion}). It is likely that the classic TI is a necessary but not sufficient ingredient of episodic accretion; additional physics must play a significant role. 

\section{Acknowledgement}

SN acknowledges the funding from the UK Science and Technologies Facilities Council, grant No. ST/S000453/1.  This research used the ALICE High Performance Computing Facility at the University of Leicester, and the DiRAC Data Intensive service at Leicester, operated by the University of Leicester IT Services, which forms part of the STFC DiRAC HPC Facility (www.dirac.ac.uk). FCSM received financial support from the European Research Council (ERC) under the European Union’s Horizon 2020 research and innovation programme (ERC Starting Grant "Chemtrip", grant agreement No 949278). AC acknowledges STFC PhD studentship funding. 

\section{Data availability}

The data obtained in our simulations can be made available on reasonable request to the corresponding author. 

\appendix
\section{Resolution study}\label{sec:Resolution_Study}

\begin{figure}
\includegraphics[width=1\columnwidth]{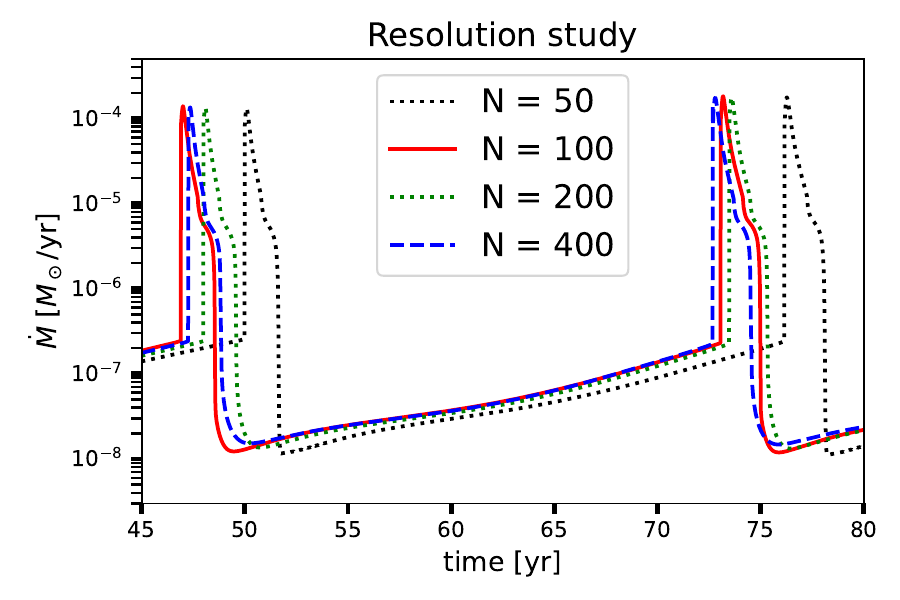}
\caption{Comparison of accretion rates on the star vs time for different values of the number of points in the radial grid, $N$. Note that for $N\ge 100$ the difference between the curves are at most a few percent, which is acceptable given other physical and modelling uncertainties in this problem.}
\label{fig:Res_study}
\end{figure}

\bref{Fig. \ref{fig:Res_study} shows a resolution study in which we kept the same problem parameters ($M_* = 0.6\msun$, $\dot M_{\rm feed} = 2\times 10^{-6}$\MSunPerYear, $\alpha_{\rm c} = 0.02$, $\alpha_{\rm h} = 0.1$) but varied the number of points in the computation grid from $N=50$ to $N=400$. The minimum and maximum radii in the grid are $R_{\rm in} = 0.015$~au and $R_{\rm out}=1$~au, respectively. We see that for $N \ge 100$, there is very little difference in the time profile of $\dot M$ on the star. This result may be surprising considering what is found in DNe TI simulations, with, e.g., \cite{Hameury98-alpha} using a default resolution of 800 radial zones. However, there is a significant difference in the disc aspect ratio in these systems. Typical radii where the instability occurs are $\sim 2$ orders of magnitude smaller in DNe compared with protoplanetary discs. Thus, $H/R = c_s/v_K$, where $v_K$ is the local Keplerian speed, is an order of magnitude smaller in DNe compared to $H/R$ for YSO (cf. Fig. \ref{fig:Classic_TI_Sigma_T}). As $H$ sets the pressure scale height in the disc, the resolution requirement is that the radial grid cell size, $\Delta R$ must be smaller than $H$ by some factor. This implies $\Delta R/R \lesssim H/R$, and thus a larger $\Delta R/R$ is possible for protoplanetary discs, requiring fewer resolution elements.}

\bibliographystyle{mnras}
\bibliography{TIP}

\bsp	
\label{lastpage}
\end{document}